\documentclass{elsart}
\usepackage{graphicx}
\usepackage{amssymb}
\begin{document}

\begin{frontmatter}

\title{Propagation of Tau Neutrinos and Tau Leptons through the Earth and 
their Detection in Underwater/Ice Neutrino Telescopes}

\author[label1]{Edgar Bugaev},
\author[label2,label3]{Teresa Montaruli},
\author[label1]{Yuri Shlepin},
\author[label3,label1]{Igor Sokalski\corauthref{cor1}}
\ead{Igor.Sokalski@ba.infn.it}
\corauth[cor1]{Corresponding author. Tel. +39-080-544-3225; 
fax: +39-080-544-2470.}
\address[label1]{Institute for Nuclear Research, 60th October Anniversary 
Prospect 7a, 117312, Moscow, Russia}
\address[label2]{Physics Department, Bari University, Via Amendola 173, 
I-70126 Bari, Italy}
\address[label3]{Istituto Nazionale di Fisica Nucleare / Sezione di Bari, Via 
Amendola 173, I-70126 Bari, Italy}

\begin{abstract}
If muon neutrinos produced in cosmological sources oscillate, neutrino 
telescopes can have a chance to detect $\tau$-neutrinos. In contrast to 
$\nu_{\mu}$'s the Earth is completely transparent for $\nu_{\tau}$'s thanks to
the short life time of $\tau$-leptons that are produced in $\nu_{\tau}\,N$
charged current interactions. $\tau$-lepton decays in flight producing another 
$\nu_{\tau}$ (regeneration chain). Thus, $\nu_{\tau}$'s cross the Earth 
without being absorbed, though loosing energy both in regeneration processes 
and in neutral current interactions. Neutrinos of all flavors can be detected 
in deep underwater/ice detectors by means of \u{C}erenkov light emitted by 
charged leptons produced in $\nu$ interactions. Muon and $\tau$-leptons have 
different energy loss features, which provide opportunities to identify 
 $\tau$-events among the multitude of muons. Some signatures of $\tau$-leptons 
that can be firmly established and are background free have been proposed in 
literature, such as 'double bang' events. In this paper we present results of 
Monte Carlo simulations of $\tau$-neutrino propagation through the Earth 
accounting for neutrino interactions, $\tau$ energy losses and $\tau$ decays. 
Parameterizations for hard part and corrections to the soft part of the 
photonuclear cross-section (which contributes a major part to $\tau$ energy 
losses) are presented. Different methods of $\tau$-lepton identification in 
large underwater/ice neutrino telescopes are discussed. Finally, we present a 
calculation of $\nu_{\tau}$ double bang event rates in km$^3$ scale detectors.
\end{abstract}

\end{frontmatter}

\section{Introduction}
\label{sec:intro}

$\nu_{\mu} \leftrightarrow \nu_{\tau}$ oscillations should lead to the 
proportion 
$\phi_{\nu_{e}}\!:\!\phi_{\nu_{\mu}}\!:\!\phi_{\nu_{\tau}}\!=\!1\!:\!1\!:\!1$ 
for neutrinos produced in cosmological sources that reach the Earth, though 
the flavor ratio at production in typical sources is expected to be 
$\phi_{\nu_{e}}\!:\!\phi_{\nu_{\mu}}\!:\!\phi_{\nu_{\tau}}\!=\!1\!:\!2\!:\!0$. 
Identification of UHE/EHE $\tau$-events in deep underwater/ice \u{C}erenkov 
neutrino telescopes (UNTs) \cite{amanda,antares,baikal,icecube,nemo,nestor} 
would confirm oscillations already discovered at lower energies 
\cite{osc1,osc2,osc3}. Measuring the ratio between cosmological 
$\nu_{\mu}$ and $\nu_{\tau}$ fluxes one could also exclude or confirm some 
more exotic scenario \cite{notau1,notau2}, such as $\nu$ decay, in which 
$\phi_{\nu_{\tau}} \ne \phi_{\nu_{\mu}}$. 

At energies $E_{\nu} \lesssim$\,1\,PeV a general approach to discriminate rare
neutrino events from the huge amount of atmospheric muons present also at 
kilometer water/ice depths is to select events from the lower hemisphere. 
These can be produced by neutrinos, the only known particle that can pass
through the Earth with negligible absorption below these energies. 
Nevertheless, $\nu$ cross-sections increase with energy. For muon neutrinos,
absorption is considerable above 1 PeV, depending on the $\nu$ zenith angle, 
hence on the path-length transversed in the Earth. On the other hand, 
$\nu_{\tau}$'s generate $\tau$-leptons via charged current (CC) interactions 
$\nu_{\tau}\,N \stackrel{CC}{\rightarrow} \tau \,N$ in the Earth. Being a 
short-lived particle, $\tau$ decays in flight producing another $\nu_{\tau}$ 
and (in $\approx 35\%$ of the cases) secondary muon and electron neutrinos 
which are generated in decay modes $\tau\rightarrow e\,\nu_{e}\nu_{\tau}$ 
($B_{e}$=17.84\%) and $\tau\rightarrow\mu\,\nu_{\mu}\nu_{\tau}$ 
($B_{\mu}$=17.36\%) \cite{PDG}. Thus, neutrinos of all flavors undergo a 
regeneration process and the Earth is transparent for $\nu_{\tau}$'s of any
energy. Nevertheless, they loose energy through the Earth due to 
$\nu_{\tau}\,N$ interactions (CC and NC) where the hadronic showers 
take away part of the energy. A minor part of the energy is lost also due to 
$\tau$-lepton propagation before decay. Thus, calculations for all flavor 
neutrino fluxes at detector location that result from propagation of an 
initial $\nu_{\tau}$ flux through the Earth must necessarily take into account
neutrino interaction properties, $\tau$-lepton energy losses and $\tau$-lepton
decay. Such calculations are reported, e.g., in 
\cite{kwi,dutta1,beacom,dutta2,yosh,bottai,giesel,sar1}, 
both for monoenergetic 
$\nu_{\tau}$ beams and for a variety of models for cosmological neutrino 
spectra. 

$\tau$-neutrinos can be detected in UNTs identifying $\tau$ leptons. To our 
knowledge the possibility of discriminating $\tau$'s from muons through their 
different energy loss properties has not yet been analyzed. In this paper we 
discuss energy loss properties and specific $\tau$-induced signatures, such 
as 'double bang' (DB) or 'lollipop' \cite{db} events, which should be affected
by negligible muon background. In \cite{athar} the results of a calculation of
the number of DB events in an UNT are published, but authors considered only 
down-going $\nu_{\tau}$'s and ignored those passing through the Earth to the 
detector from the lower hemisphere. In \cite{dutta1,dutta3,beacom1} 
the measurement of the ratio of up-going shower-like and track-like
events is proposed,
instead of an event-by-event identification of $\tau$- or $\mu$-events. As a 
matter of fact, $\tau$-leptons, as well as muons, produce events of both kinds
in a UNT. $\nu_{e}$'s produce shower-like events, as well. In case $\nu_{\mu}$ 
oscillate into $\nu_{\tau}$, the fraction of shower-like events is larger than
what expected for no oscillations.

In this paper we describe the results, preliminarily given in 
\cite{ourtau1,ourtau2}, of a MC simulation of $\tau$-neutrino propagation 
through the Earth. Both monoenergetic neutrino beams and spectra predicted by 
Protheroe \cite{proth} and Mannheim et al. \cite{mpr} (which were not 
discussed in \cite{dutta1,beacom}) have been considered. We have taken into 
account corrections to photonuclear interaction cross-sections which play the 
main role in $\tau$-lepton energy losses, compared to pair production and 
bremsstrahlung.  We used results published in \cite{bugaev1} where {\it i)} 
the soft part of photonuclear cross-section based on the generalized vector 
dominance model originally published in \cite{pbb1,pbb2} is corrected and 
{\it ii)} the hard part is developed in the QCD perturbative framework. The 
soft part corrections to photonuclear interaction, not accounted for in 
\cite{kwi,dutta1,beacom,dutta2,yosh,bottai,giesel,sar1}, 
have lead to an increase 
of the total $\tau$-lepton energy losses by 20--30\% in UHE/EHE range. Also we 
have analyzed the possibility to distinguish UHE/EHE $\tau$-leptons from 
muons in a UNT by different energy losses along tracks through production of 
electromagnetic and hadronic showers generated in bremsstrahlung, 
$e^{+}e^{-}$-pair production and photonuclear interaction. 

In Sec.~\ref{sec:method} we describe the the tools used for this calculation 
with particular attention to $\tau$-lepton energy losses for photonuclear 
interaction; in Sec.~\ref{sec:results} results on propagation of UHE/EHE 
$\tau$-neutrinos through the Earth are reported. In Sec.~\ref{sec:sign} we 
analyze the possibility to distinguish $\tau$-lepton from muon in a UNT thanks
to their different energy loss properties and we give results on a calculation
of the rate of DB events coming from both hemispheres in km$^3$ scale UNTs; in 
Sec.~\ref{sec:conclusions} we conclude. In Appendix~\ref{app:PN} the corrected
formula for muon and $\tau$-lepton photonuclear cross-section is given. 

\section{Tools for $\nu_{\tau}$ propagation in the Earth}
\label{sec:method}

A MC calculation has been developed to account for the following processes:
\begin{itemize}
\item
NC neutrino interactions that cause neutrino energy losses.
\item
CC interactions. For $\nu_{e}$'s and $\nu_{\mu}$'s it has been assumed that 
resulting electrons and muons are absorbed, while for $\nu_{\tau}$'s 
$\tau$-leptons are followed.
\item
$\tau$-lepton propagation through the Earth accounting for its energy losses.
\item
$\tau$-lepton decay resulting in another $\nu_{\tau}$ appearance and in 
$\approx$35\% of the cases also in $\nu_{e}$ or $\nu_{\mu}$ production. 
$\nu_{\tau}$'s generated in $\tau$ decays have been reprocessed again through 
this chain of processes. Thus, the 'regeneration chain' 
$\nu_{\tau} \rightarrow \tau \rightarrow \nu_{\tau} \rightarrow \cdots$ has 
been simulated. 
\end{itemize}
\begin{figure}[htb]
\begin{center}
\includegraphics[height=12.5cm]{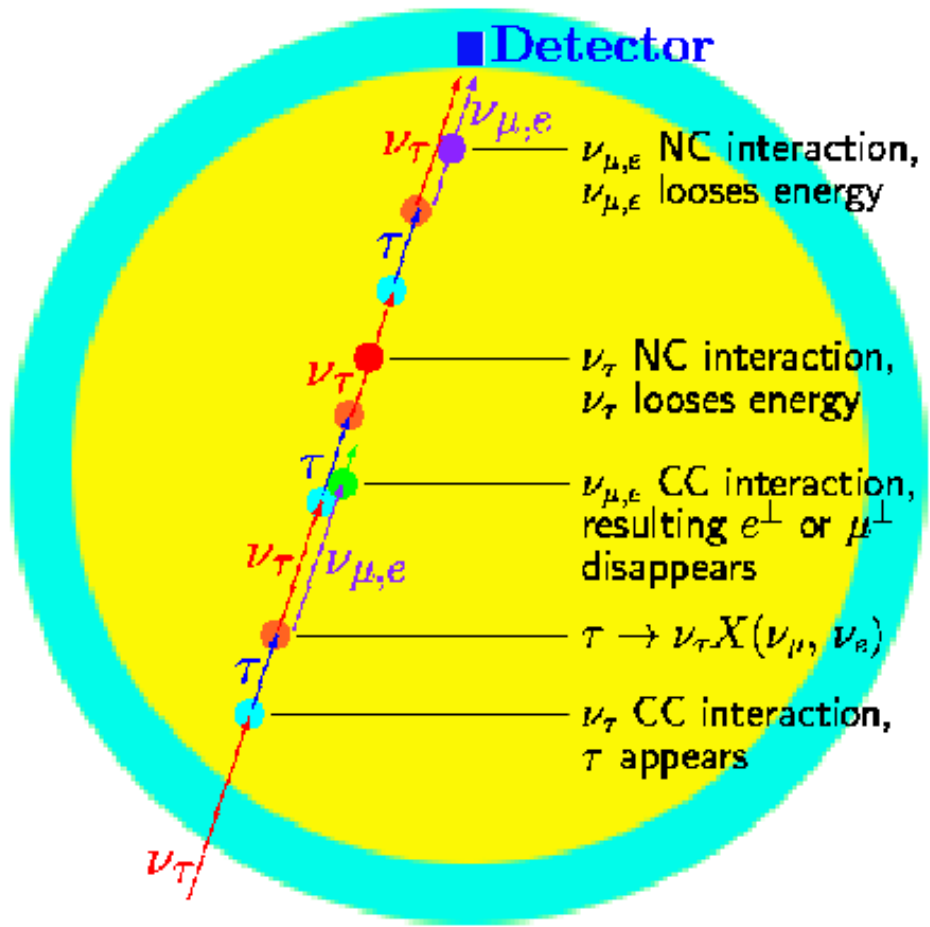}
\end{center}
\caption{\label{f1}
Schematic view of processes induced by high energy $\nu_{\tau}$'s when 
propagating through the Earth.}
\end{figure}
We have assumed the Earth composition made by standard rock ($A$=22, $Z$=11) 
of variable density with the Earth density profile published in \cite{earth}.
Processes undergone by $\tau$-neutrinos during propagation through the Earth 
are shown schematically in Fig.~\ref{f1}.

\subsection{Neutrino interactions}
\label{sec:interact}

\begin{figure}[htb]
\begin{center}
\includegraphics[height=13.9cm]{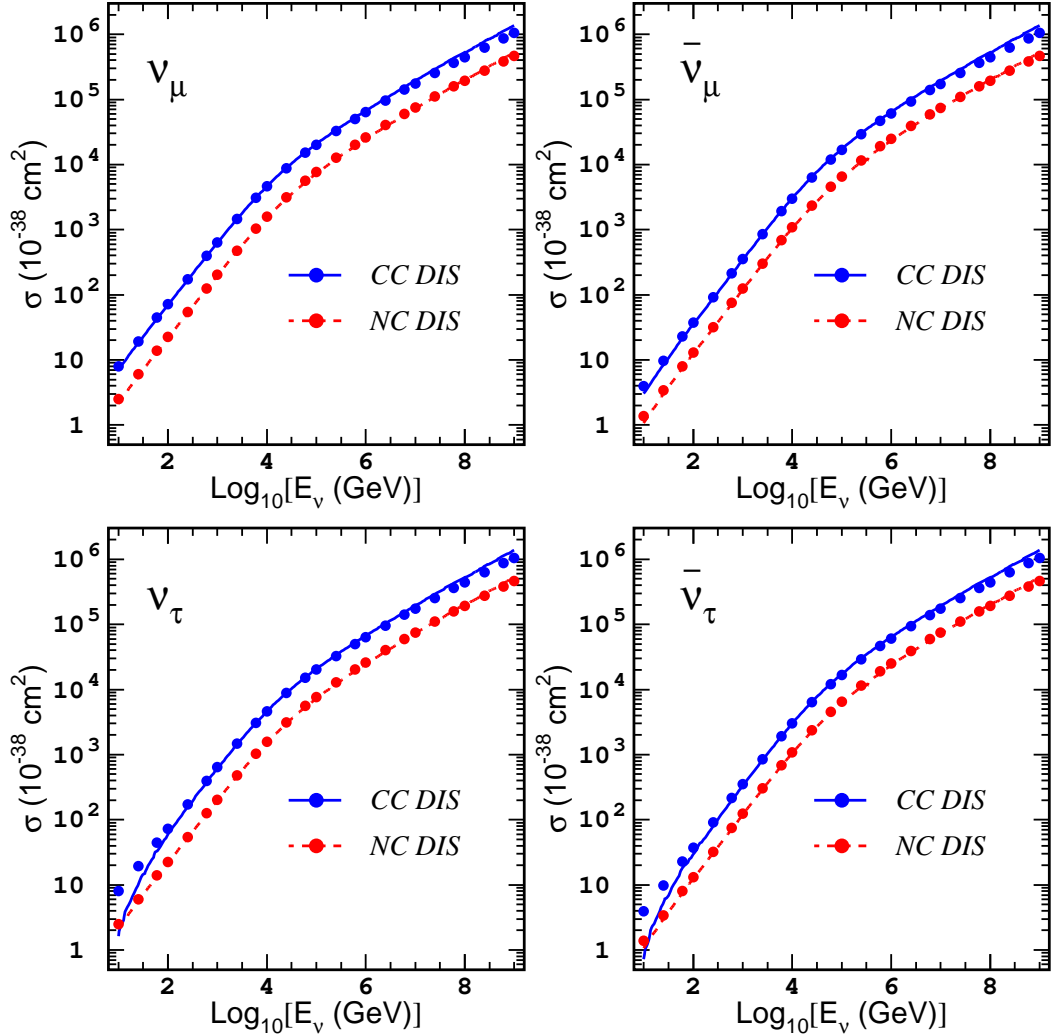}
\end{center}
\caption{\label{f2}
CC (solid lines) and NC (dashed lines) deep inelastic scattering 
cross-sections for $\nu_{\tau}$, $\bar \nu_{\tau}$, $\nu_{\mu}$, and 
$\bar \nu_{\mu}$ (cross-sections for electron (anti-)neutrinos practically 
coincide with ones for muon (anti-)neutrinos) for isoscalar medium with 
$Z/A$\,=\,0.5. Lines: cross-sections with structure functions CTEQ3-DIS (used 
in this work); markers: cross-sections with CTEQ4-DIS (from 
\protect\cite{CTEQ4GAN}). 
}
\end{figure}

To simulate $\nu\,N$ interactions we have used a generator developed by 
P.~Lipari and F.~Sartogo which is based on cross-sections described in 
\cite{lip1}. It has been adapted for our use up to the high energy region we 
are interested in ($E_{\nu} > 10^5$~GeV) adjusting the efficiency of the
rejection technique method described in \cite{LEPTO}. Both for NC and CC 
interactions we have accounted only for deep inelastic scattering since in the 
considered energy range other channels contribute negligibly. For what 
concerns deep inelastic scattering, the generator uses the electroweak 
standard formula for the inclusive differential cross-section and it is based 
on the LUND packages for hadronization \cite{LEPTO,hadronization,JETSET}. 
Structure functions CTEQ3-DIS \cite{CTEQ} taken from PDFLIB \cite{PDFLIB} 
suited for the high energy regime have been used \footnote{It should be 
noticed, however, that the CTEQ3 parton function extrapolation at small 
Bjorken $x_{Bj}$, as implemented in PDFLIB, is more similar to the CTEQ6 one 
rather than to the CTEQ4 and CTEQ5.}. More recent versions of CTEQ PDFs 
\cite{CTEQ4,CTEQ5,CTEQ6} produce differences on cross-sections at the level of
up to 25\% at $10^9$\,GeV (see Fig.~\ref{f2}). 

\subsection{Tau-lepton energy losses}
\label{sec:losses}

The MUM package (version 1.5) has been used to simulate $\tau$-lepton 
propagation through matter. Comparing to \cite{mum} (where the first
version of the package originally developed for the muon propagation was 
described) the package has been extended to treat $\tau$-leptons, accounting 
for their short life time and large probability to decay in flight. Besides, 
the newest corrections to photonuclear cross-section have been introduced. 
Formulas for cross-sections for $e^{+}e^{-}$-pair production, bremsstrahlung, 
knock-on electron production and stopping formula for ionization implemented 
in MUM\,1.5 can be found in \cite{mum} (Appendix A)~\footnote{When dealing 
with $\tau$-lepton propagation one must change muon mass for $\tau$-lepton 
mass in all the formulas except for expression for 
$q_c = 1.9\,m_{\mu} / Z^{1/3}$ in formula for bremsstrahlung cross-section 
(see page 074015-15, Appendix A in \cite{mum}) where the muon mass was 
introduced just as a mass-dimension scale factor.} where they are given 
according to \cite{brembb1,brembb2,pk1,pk2,pk3,pk4,pk5,lohman,ion1,ion2,ion3}.

Main improvements in MUM\,1.5 concern photonuclear interaction of leptons with
nucleons, in which virtual photons are exchanged. We are interested in the 
diffractive region of the kinematic variable space ($q^2<0$, transferred 
energy $\nu$ large, $x_{Bj}\sim q^2/2 \nu$ small). The photonuclear 
interaction has been treated as the absorption of the virtual photon by the 
nucleon and, using the optical theorem, it has been connected with the Compton
scattering of a virtual photon, $\gamma^*+N \to \gamma^*+N$. The Compton 
scattering in the diffractive region has been described by the vacuum exchange
which, in turn, is modeled in QCD by the exchange of two or more gluons in a 
color singlet state. This is possible because, in the laboratory system, the 
interaction region has a large longitudinal size, and the photon develops an 
internal structure due to its coupling with quark fields. In the diffractive 
region, $\gamma^*N$-scattering dominates the Compton amplitude, while the 
contribution to it due to the photon bare component is smaller.

It has been shown by HERA experiments 
\cite{HERA1,HERA2,HERA3,HERA4,HERA5,HERA6} that the picture of the usual soft 
diffraction, soft pomeron exchange, does not work well if the center-of-mass 
energy of $\gamma^*$ and the target nucleon $N$ is very large (i.e. if $q^2$ 
and $\nu$ are in the diffractive region). For a description of the data in the
framework of the simplest Regge-pole model, one needs at least two pomerons: 
soft and hard ones. A two-component picture of the photonuclear interaction in
the diffractive region arises very naturally due to inherent QCD properties 
(asymptotic freedom, confinement, color transparency) and is the common 
feature of some recent quantitative models \cite{bugaev1,gotsman,martin}. 
Generally, the photonuclear cross-section integrated over $Q^2 = -q^2$ is 
expressed through the electromagnetic structure functions of nuclei 
$\sigma_{L,T}$ by the formula
\begin{equation}
\label{sigmabb1}
\frac{d\sigma^{pn}}{dv}=\frac{\alpha}{2\pi v}\int\limits^{\infty}_{Q^2_{min}} \frac{dQ^2}{Q^2} \{v^2\sigma_T(1-\frac{2m^2_l}{Q^2})+2(1-v)(\sigma_T+\sigma_L)\},
\end{equation}
where
$Q^2_{min}\cong m^2_lv^2 / (1-v)$, $v= \Delta E / E$ is the fraction of energy
lost in the interaction by a lepton of energy $E$ with mass $m_l$ and 
$\alpha =$\,1/137 is the fine structure constant.

In the first version of the MUM code \cite{mum} only the soft part 
of this integral was accounted for according to Bezrukov-Bugaev (BB) 
parameterization that was developed in the frame of the generalized vector 
dominance model \cite{pbb1,pbb2}. In MUM\,1.5 two main improvements have been 
done. Firstly, new terms have been  included in the BB parameterization 
\cite{bugaev1}. Since these produce negligible effects in the case of muon, 
they were not introduced in the original formula given in \cite{pbb1,pbb2}. 
Nevertheless, these terms are essential in case of $\tau$-lepton (due to its 
larger mass compared to muon: $m_{\mu}$=0.1057\,GeV; $m_{\tau}$=1.777\,GeV) 
and they increase total $\tau$-lepton energy losses by $\sim$20$\div$30\% in 
UHE/EHE range. Secondly, the hard component of the photonuclear cross-section 
has been introduced using results obtained in \cite{bugaev1}. This part is 
described by the phenomenological formula based on the color dipole model 
\cite{mueller,nikolaev}. The main element of the approach used in 
\cite{bugaev1} is the total cross-section $\sigma(r_{\bot},s)$ for scattering 
of dipoles ($q\bar{q}$-pairs) of a given transverse size $r_{\bot}$ from a 
target proton at fixed center-of-mass energy squared $s$. The 
$r_{\bot}$-dependence of $\sigma(r_{\bot},s)$ is qualitatively predicted by 
perturbative QCD:
\begin{equation}
\label{sigmabb3}
\sigma(r_{\bot},s)\sim
r^2_{\bot}e^{-\frac{r_{\bot}}{r^0_{\bot}}}(r^2_{\bot}s)^{\lambda_H}.
\end{equation}
In this formula $\lambda_H$ and $r^0_{\bot}$ are parameters determined from a
comparison of the total photonuclear cross-section (or structure functions) 
with HERA data at small $x_{Bj}$. Such an expression for $\sigma (r_{\bot},s)$
was suggested in \cite{forshaw}. Besides, for a region of large $s$ (outside 
of the HERA region) the unitarization procedure is used in \cite{bugaev1}, 
which leads, asymptotically, to a logarithmic $s$-dependence of the dipole 
cross-section,
\begin{equation}
\label{sigmabb4}
\sigma^{unit}(r_{\bot},s)\sim \lambda_H \ln s.
\end{equation}
The hard part of the photonuclear cross-section is shown in Fig.4 and Fig.5 in
\cite{bugaev1} for muons and $\tau$'s, respectively. We have made a polynomial
parameterization for this part that is given in Appendix \ref{app:PN} along 
with the complete formula for photonuclear cross-section used in this work.

\begin{figure}[htb]
\begin{center}
\includegraphics[height=13.4cm]{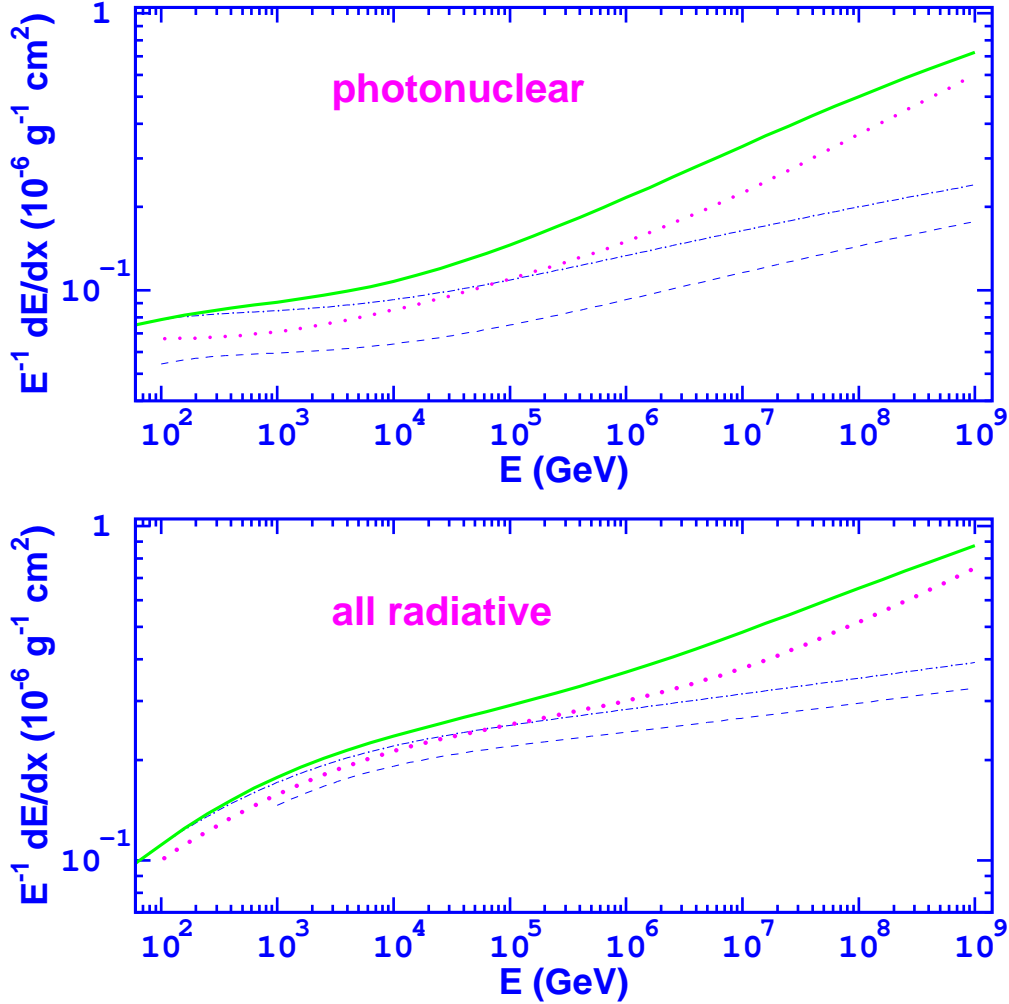}
\end{center}
\caption{\label{f3}
$\tau$-lepton energy losses in standard rock. Upper plot: energy losses due to
photonuclear interaction. Lower plot: energy losses due to all radiative 
processes, including bremsstrahlung and direct $e^{+}e^{-}$-pair production 
computed according to \cite{brembb1,brembb2,pk1,pk2,pk3,pk4,pk5}. In each 
plot: dashed line is for non corrected soft part \cite{pbb1,pbb2}; dash-dotted
line for corrected soft part \cite{bugaev1,pbb1,pbb2}; dotted line includes 
hard component of photonuclear cross-section according to \cite{duttaloss}; 
solid line: includes hard component according to \cite{bugaev1} (as it is 
treated in this work). 
}
\end{figure}

$\tau$-lepton energy losses in standard rock ($A$=22, $Z$=11) for photonuclear
interaction and for the sum of all radiative  processes are presented in 
Fig.~\ref{f3}. The importance of both corrections described above compared
to the soft BB formula can be clearly seen. Results published in 
\cite{duttaloss} for $\tau$-lepton energy losses including another calculation
of the hard component of the photonuclear cross-section are shown for 
comparison, as well. The main difference between \cite{bugaev1} and 
\cite{duttaloss} is due to the fact that corrections to the BB formula (soft 
part) \cite{pbb1,pbb2} have not been introduced in \cite{duttaloss}. The hard 
component of the photonuclear cross-section differs in \cite{bugaev1} and
\cite{duttaloss}, as well, since theoretical approaches applied in these two 
works are completely different.
 
\subsection{Tau-lepton decay}
\label{sec:decay}

To generate $\tau$-lepton decays we used the TAUOLA package \cite{tauola} 
which was developed for SLC/LEP experiments where $\tau$-leptons are produced 
in $e^{+}e^{-}$ collisions: 
\mbox{$e^{+}e^{-} \rightarrow Z \rightarrow \tau^{+}\tau^{-}$}. TAUOLA 
generates decays of $\tau^{\pm}$'s of a given energy, taking into account all 
the effects of $\tau$ spin polarization. 22 decay modes are treated (sum to 
almost 100\% of the total width) including modes that are responsible for 
$\nu_{e}$ and $\nu_{\mu}$ appearance: 

\begin{equation}
\label{decay1}
\tau^{-} \rightarrow \,\, e^{-} \,\, \bar \nu_{e} \,\, \nu_{\tau}\,\,\,\,\,\,\,
(\tau^{+} \rightarrow \,\, e^{+} \,\, \nu_{e} \,\, \bar \nu_{\tau})
\end{equation}
\begin{equation}
\label{decay3}
\tau^{-} \rightarrow \,\, \mu^{-} \,\, \bar \nu_{\mu} \,\, \nu_{\tau}\,\,\,\,\,\,\,
(\tau^{+} \rightarrow \,\, \mu^{+} \,\, \nu_{\mu} \,\, \bar \nu_{\tau})
\end{equation}

In our simulation we tracked only neutrinos (of all flavors) resulting from 
$\tau$ decay, since ranges of charged leptons are negligibly small compared to
the Earth dimensions. Decay lengths of $K$'s and $\pi$'s resulting from $\tau$
decay are much longer than their interaction lengths in the considered energy 
range, hence we did not simulate their decay neglecting secondary 
neutrinos~\footnote{By the same reason we did not follow $K$'s and $\pi$'s 
produced in $\nu$ CC interactions.}.

\section{Results on $\nu_{\tau}$ propagation through the Earth}
\label{sec:results}

In this section results on the simulation of propagation of $\nu_{\tau}$ and 
$\bar \nu_{\tau}$ monoenergetic beams through the Earth are presented where 
all processes shown in Fig.~\ref{f1} have been accounted for.

\subsection{Results on monoenergetic tau-neutrino beams}
\label{sec:mono}

\begin{figure}[htb]
\begin{center}
\includegraphics[width=13.8cm]{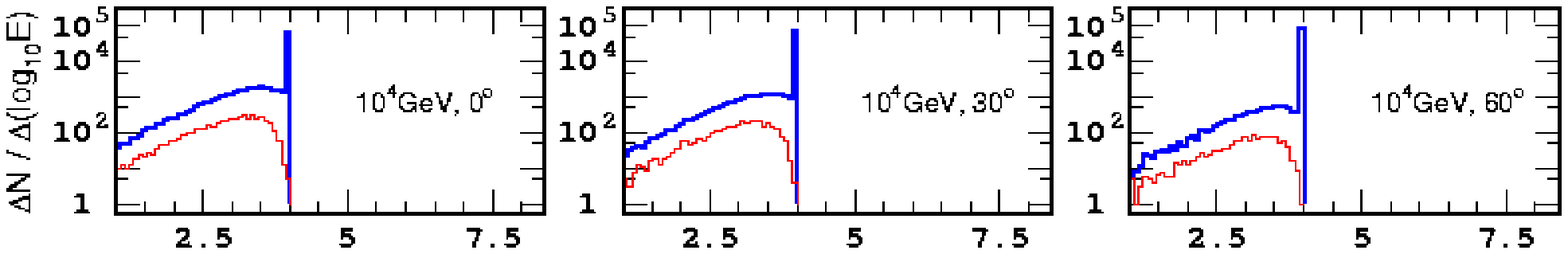}
\vspace{-8mm}
\includegraphics[width=13.8cm]{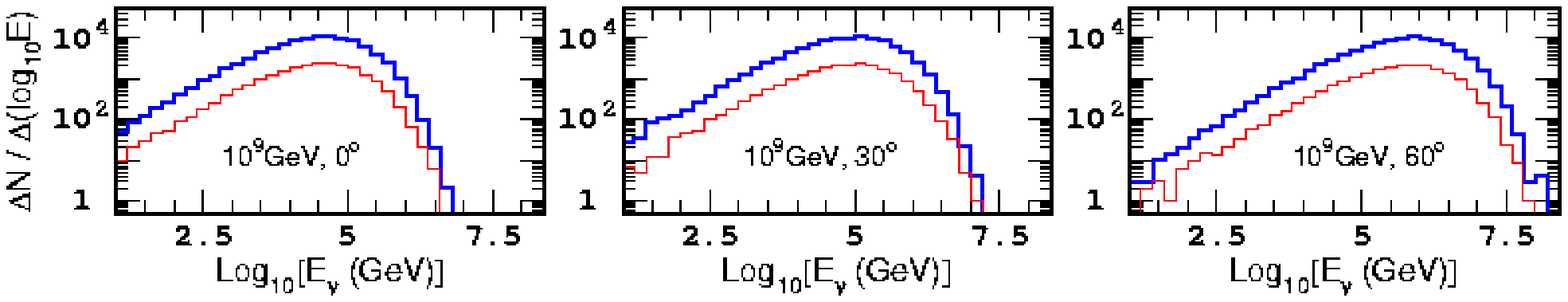}
\end{center}
\vspace{5mm}
\caption{\label{f4}
Monoenergetic $\nu_{\tau}$ beams incident on the Earth at nadir angles 
0$^{\circ}$ (left), 30$^{\circ}$ (center), 60$^{\circ}$ (right). Results after 
propagation through the Earth. Upper row: $E^{0}_{\nu}=$10 TeV, lower row: 
$E^{0}_{\nu}=$ 1 EeV. Upper spectra: $\nu_{\tau}$; lower spectra: secondary 
$\nu_{\mu}$ ($= \nu_{e}$).
}
\end{figure}

\begin{figure}[htb]
\begin{center}
\includegraphics[width=10.7cm]{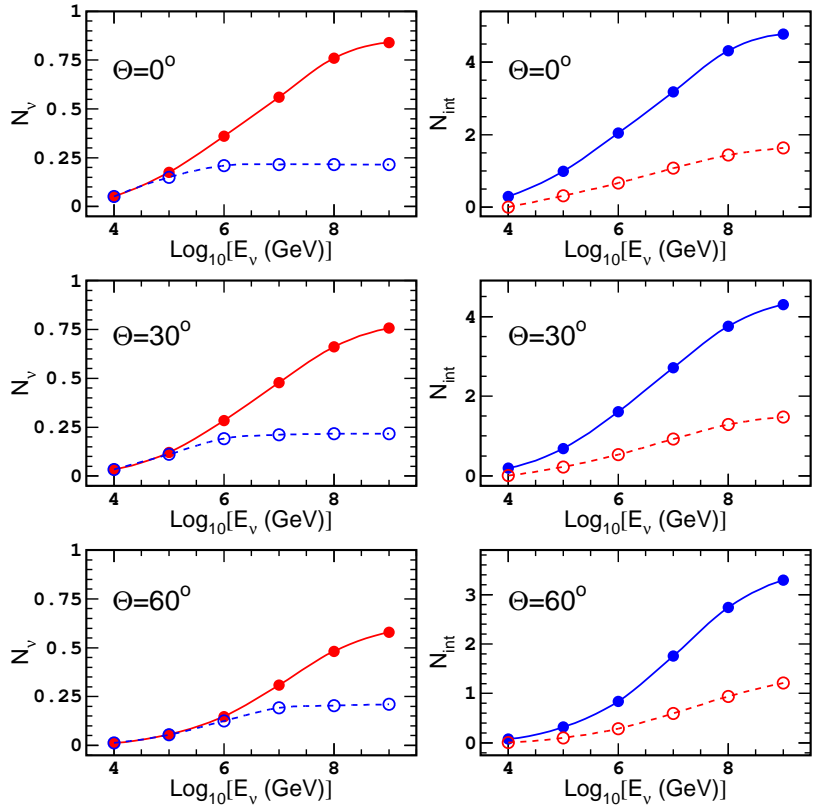}
\end{center}
\caption{\label{f5_6}
Left column: Mean number of secondary $\nu_{\mu}$'s ($= \nu_{e}$'s) per 
$\nu_{\tau}$ that are generated when a $\nu_{\tau}$ crosses the Earth 
undergoing regeneration processes vs $E_{\nu_{\tau}}$ (solid lines); mean 
number of secondary $\nu_{\mu}$'s ($= \nu_{e}$'s) that survive and come out 
from the Earth (dashed lines). Right column: mean number of CC (solid lines) 
and NC (dashed lines) interactions that occurs in the chain 
$\nu_{\tau}\rightarrow\tau\rightarrow\nu_{\tau}\cdots$ when a $\tau$-neutrino 
crosses the Earth vs its energy. Results for $\nu_{\tau}$ and 
$\bar \nu_{\tau}$ are identical. 
}
\end{figure}

The outgoing spectra of initially monoenergetic $\nu_{\tau}$'s after 
propagation through the Earth for three nadir angles of incidence on the Earth
are shown in Fig.~\ref{f4} for two neutrino energies: 10 TeV and 1 EeV. 
Moreover, the outgoing spectra of secondary $\nu_{\mu}$'s and $\nu_{e}$'s 
that are produced in decays of $\tau$-leptons generated in $\nu_{\tau}$ CC 
interactions are presented. Results for $\nu_{\tau}$ and $\bar \nu_{\tau}$ 
beams are identical. At $E_{\nu_{\tau}}=$10$^{4}$\,GeV neutrino cross-sections
are small and most of $\tau$-neutrinos cross the Earth without interacting 
keeping their initial energy (see the peaks in the upper panels in 
Fig.~\ref{f4}). The fraction of such neutrinos is lower at the nadir direction
($\theta=$\,0$^{\circ}$) since 
the amount of matter crossed is maximum. This fraction increases with nadir 
angle. At $E_{\nu_{\tau}} = 10^{9}$\,GeV all $\tau$-neutrinos undergo at least
one interaction and the peak at initial energy disappears since neutrinos 
loose energy both in NC interactions and in regeneration chain. The fraction 
of secondary $\nu_{e}$'s and $\nu_{\mu}$'s produced by the incoming 
$\nu_{\tau}$ is 5\% at $E_{\nu_{\tau}}=$10$^{4}$\,GeV and 22\% at 
$E_{\nu_{\tau}}=$10$^{9}$\,GeV for the nadir direction~\footnote{We consider 
all secondaries collinear with respect to the primary $\nu_{\tau}$, an 
approximation that at the energies to which we are interested is reasonable.}.
These numbers are in good agreement with results published in \cite{beacom}.

\begin{figure}[htb]
\begin{center}
\includegraphics[width=10.5cm]{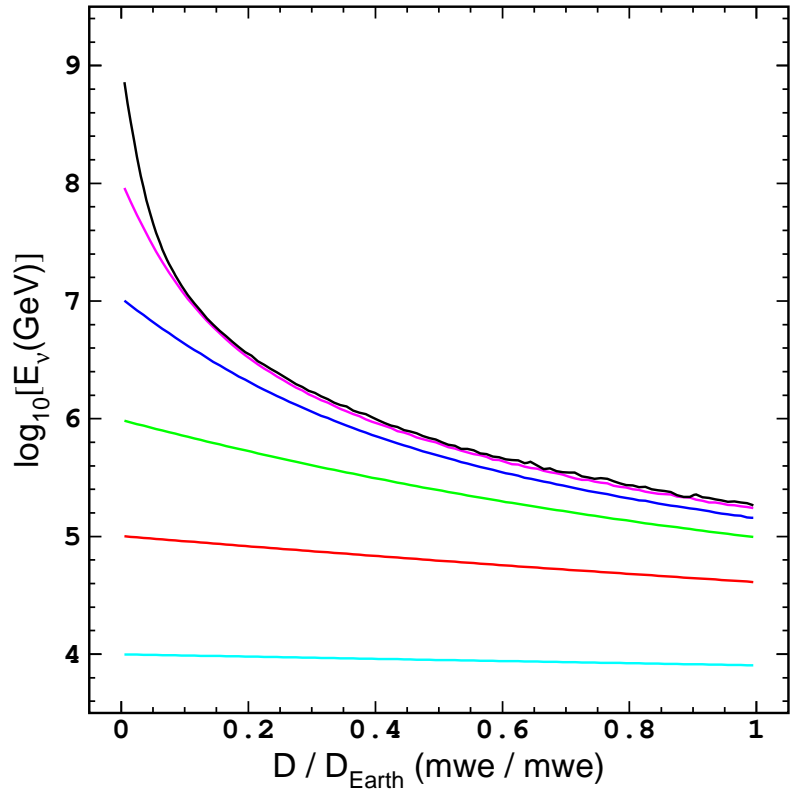}
\end{center}
\caption{\label{f7}
Mean energy degradation for $\nu_{\tau}$ with initial energies (from bottom to
top): 10$^4$\,GeV, 10$^5$\,GeV, 10$^6$\,GeV, 10$^7$\,GeV, 10$^8$\,GeV, 
10$^9$\,GeV when penetrating the Earth at nadir angle 0$^o$ vs fraction of 
Earth diameter crossed.}
\end{figure}

The fraction of secondary neutrinos that accompany $\tau$-neutrinos emerging 
from the Earth increases with initial $\tau$-neutrino energy but saturates at 
the level of 0.22 at some critical energy which depends upon nadir angle 
(Fig.~\ref{f5_6}, left column). These results are also in a good agreement 
with \cite{beacom}. The CC cross-section increases with energy and, 
consequently, the number of secondaries generated along $\tau$-neutrino path 
increases also. Nevertheless, this growth is moderated by absorption of 
secondary $\nu_{\mu}$'s and $\nu_{e}$'s which, in contrast to $\nu_{\tau}$'s, 
do not regenerate. In the right column in Fig.~\ref{f5_6} the mean number of 
NC and CC interactions occurring to a primary $\nu_{\tau}$ ($\bar \nu_{\tau}$)
when it travels through the Earth is shown The number of CC interactions 
corresponds to the number of regeneration steps. This number grows both with 
$\tau$-neutrino energy and with amount of matter crossed, which is maximum at 
the nadir.

Fig.~\ref{f7} shows $\tau$-neutrino energy losses due to regeneration process 
and NC interactions when crossing the Earth. 

\subsection{Results for some astrophysical diffuse $\nu$ spectra}
\label{sec:real}

\begin{figure}[htb]
\begin{center}
\includegraphics[width=11.9cm]{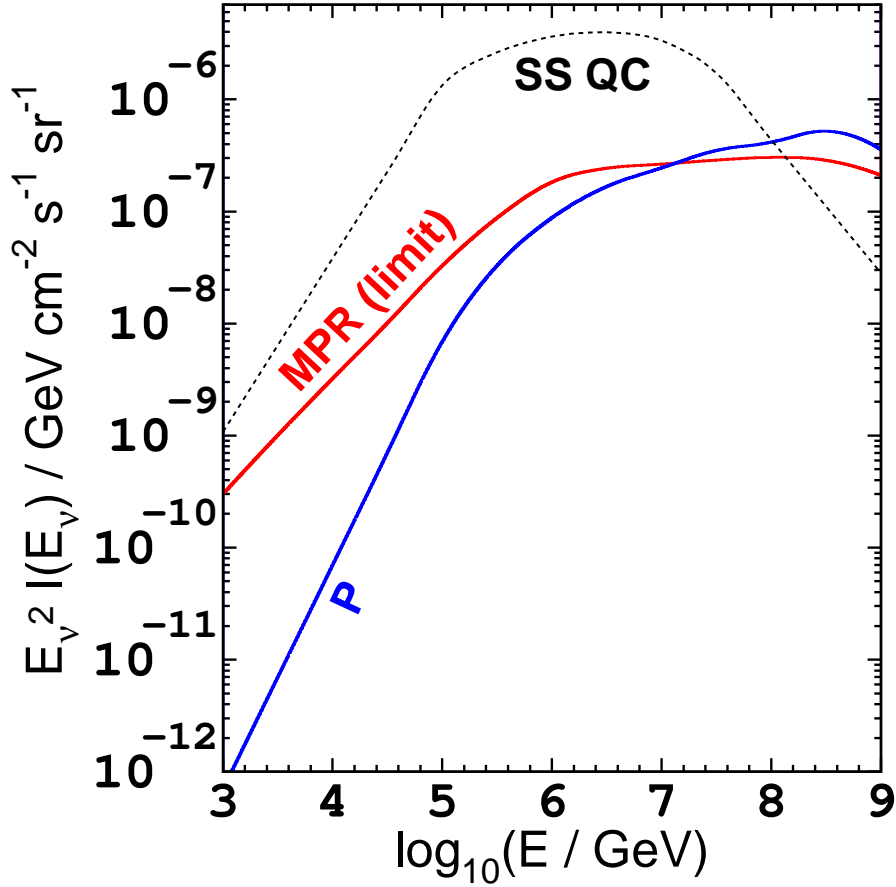}
\end{center}
\caption{\label{f8}
Protheroe diffuse neutrino spectrum \cite{proth}, MPR limit on diffuse 
neutrino spectrum \cite{mpr} and 'Quasar background flux prediction' (SS QC) 
\cite{SS}. 
}
\end{figure}

We performed MC simulations for two spectra of cosmological neutrinos: the 
spectrum predicted by Protheroe~\cite{proth} for an 
optically thick proton blazar model and an upper bound (not a source model) 
on diffuse $\nu$ spectrum for optically thick sources \cite{mpr} (MPR 
bound)~\footnote{The 'Quasar background flux prediction' \cite{SS} was not 
considered, though it is more optimistic, since it is at the exclusion level 
by AMANDA \cite{amanda1a,amanda1b} and Baikal \cite{baikal1} experiments.} 
(see Fig.~\ref{f8}).

We have considered neutrinos of all flavors in the incoming flux including 
$\nu_{e}$'s and $\nu_{\mu}$'s produced in sources and $\nu_{\tau}$'s that 
appear on the way to the Earth due to $\nu_{\mu} \leftrightarrow \nu_{\tau}$ 
oscillations. Neutrinos and anti-neutrinos of all flavors have been assumed to
be present in the proportion $\phi_{\nu}\!:\!\phi_{\bar \nu}\!=\!1\!:\!1$ in 
astrophysical neutrino fluxes. The background of $\nu_{\mu,e}$ atmospheric 
neutrinos from pion and kaon decays (conventional neutrinos), as well as the 
contribution of prompt $\nu$'s from charmed mesons (calculated in the 
Recombination Quark-Parton Model frame) was simulated using spectra taken
from \cite{naumov1}. 

\begin{figure}[htb]
\begin{center}
\includegraphics[height=14.2cm]{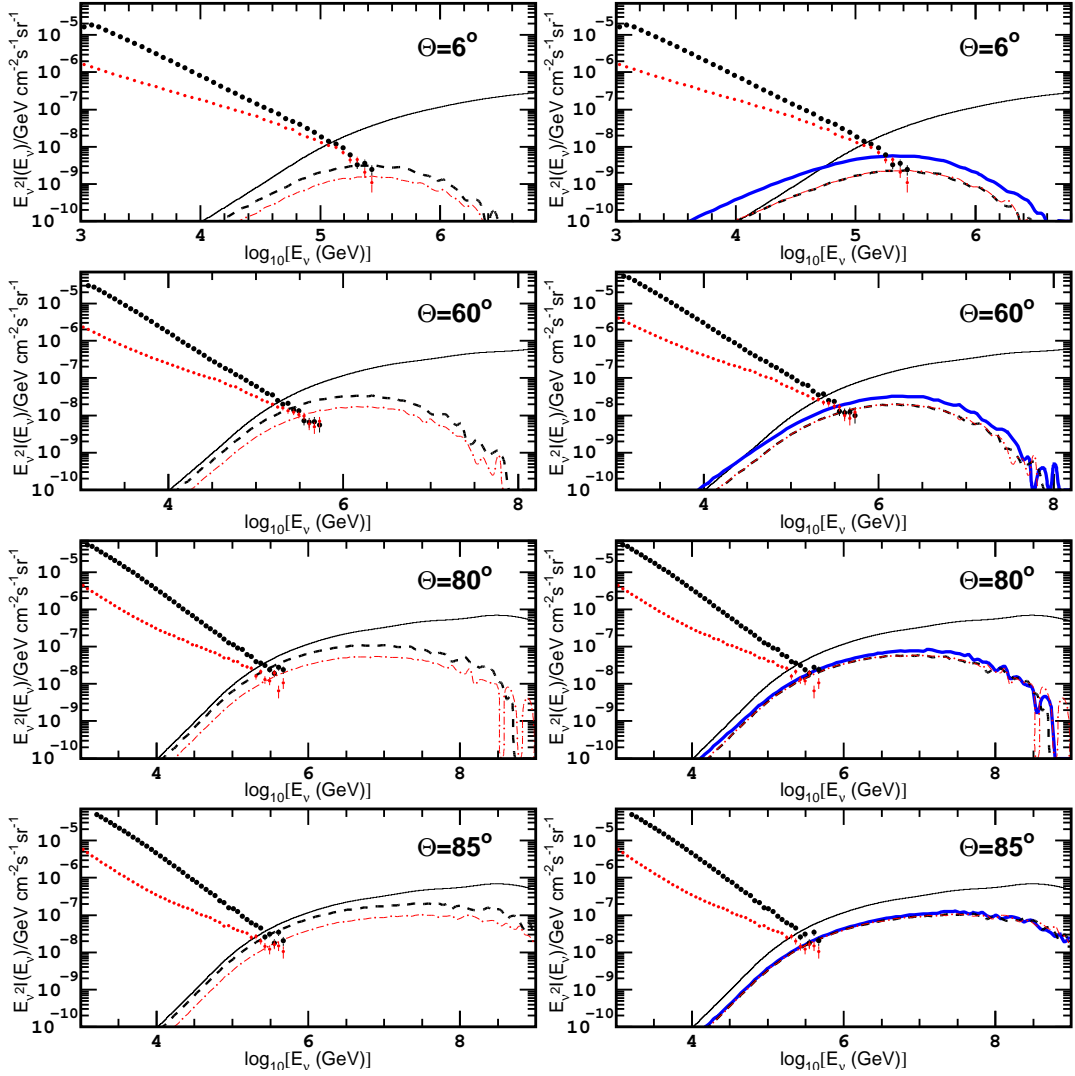}
\end{center}
\caption{\label{f9}
Spectra of cosmological $\nu$'s sampled according to Protheroe spectrum
(\protect\cite{proth}) incident on the Earth with 4 different nadir angles 
and after propagation through it. Left column: no oscillations, 
$\phi_{\nu_{e}}\!:\!\phi_{\nu_{\mu}}\!:\!\phi_{\nu_{\tau}}=
1\!:\!2\!:\!0$, 
right column: oscillations with maximal mixing, 
$\phi_{\nu_{e}}\!:\!\phi_{\nu_{\mu}}\!:\!\phi_{\nu_{\tau}}=1\!:\!1\!:\!1$. 
Thin solid lines: total incoming flux of astrophysical $\nu$'s; thick solid 
(right panels), dashed and dash-dotted lines (overlapping in right panels): 
out-coming 
$\nu_{\tau} + \bar \nu_{\tau}$, $\nu_{\mu} + \bar \nu_{\mu}$, 
$\nu_{e} + \bar \nu_{e}$ after propagation through the Earth, respectively.
$\nu_{\mu,e}$ spectra include secondary neutrinos produced by $\nu_{\tau}$'s.
In the left part of each panel atmospheric neutrino spectra 
(conventional$+$prompt~\cite{naumov1}) after propagation through the Earth are
shown, upper thick markers: $\nu_{\mu} + \bar \nu_{\mu}$, lower thin markers: 
$\nu_{e} + \bar \nu_{e}$. Atmospheric neutrino spectra are given up to 
$\approx$300 TeV where cosmological neutrino fluxes start to dominate since 
the simulation statistics at higher energies is very poor due to the steepness
of spectra.
}
\end{figure}
\begin{figure}[htb]
\begin{center}
\includegraphics[height=14.2cm]{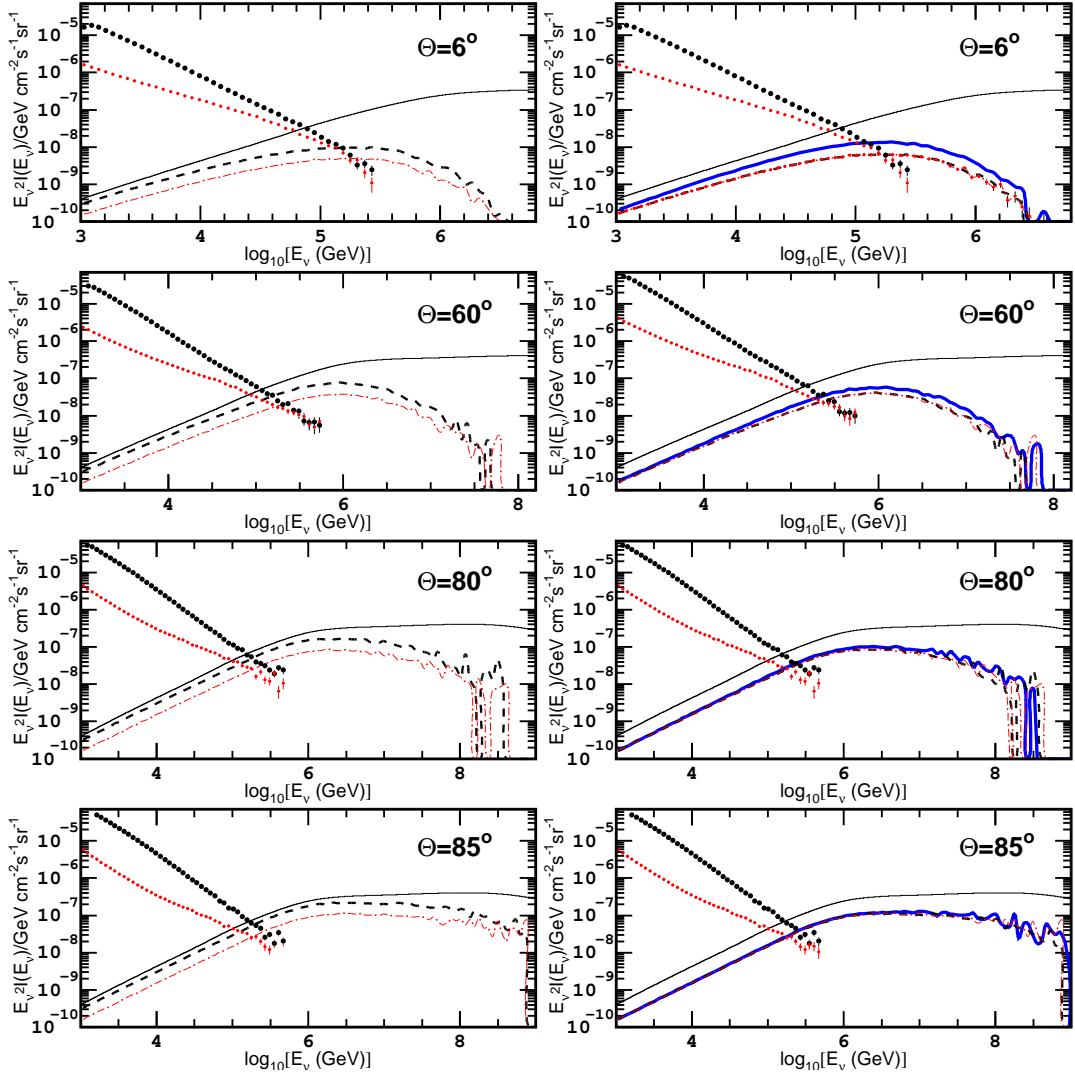}
\end{center}
\caption{\label{f10}
The same as in Fig.~\protect\ref{f9} but astrophysical neutrinos are generated
according to MPR upper limit on diffuse neutrino flux~\protect\cite{mpr}.
}
\end{figure}
\begin{figure}[htb]
\begin{center}
\includegraphics[height=14.2cm]{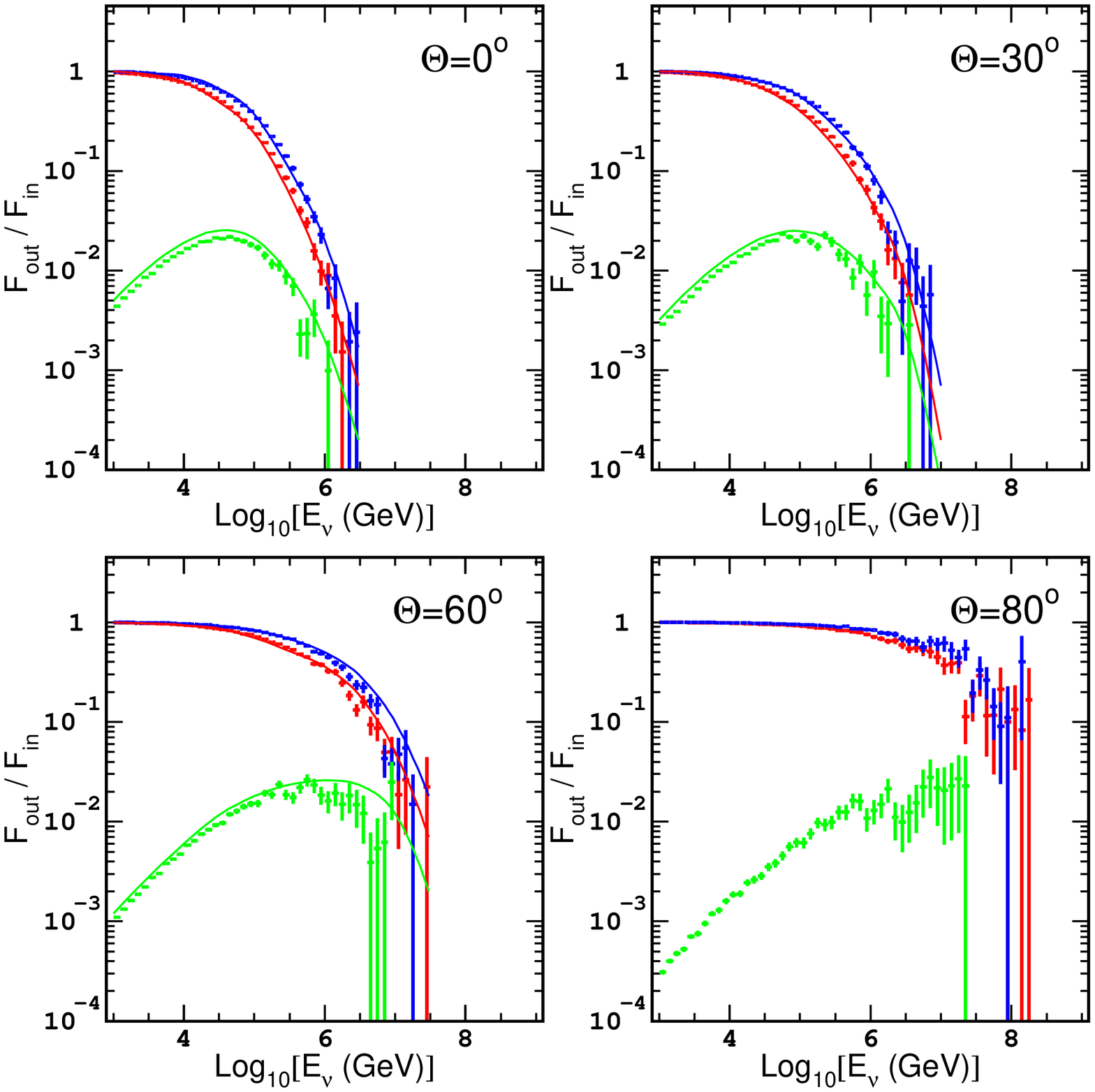}
\end{center}
\caption{\label{f11}
The ratio of $\nu_{\tau} + \bar \nu_{\tau}$ (upper curves),
 $\nu_{\mu} + \bar \nu_{\mu}$ (middle curves) fluxes after propagation through
the Earth and of secondary $\nu_{\mu} + \bar \nu_{\mu}$ 
($=\nu_{e}+\bar \nu_e$) (lower curves) over incident flux $\propto E^{-2}$ for
four nadir angles. Markers: this work, lines: results published in 
\cite{dutta2} for nadir angles $\theta$\,=0$^{\circ}$, $\theta$\,=30$^{\circ}$
and $\theta$\,=60$^{\circ}$ (data for $\theta$\,=80$^{\circ}$ are absent in
\cite{dutta2}). 
}
\end{figure}

Results for Protheroe and MPR spectra for astrophysical neutrinos incident on 
the Earth with 4 different nadir angles and transformed after propagation 
through the Earth are shown in Fig.~\ref{f9} and Fig.~\ref{f10}. 
$\tau$-neutrino fluxes exceed  $\nu_{\mu,e}$ ones remarkably up to nadir 
angles $\theta \sim 60^o$ since $\nu_{\tau}$'s are not absorbed by the Earth 
in contrast to $\nu_{\mu,e}$'s. But their spectra are shifted to lower 
energies with respect to initial spectra due to energy degradation in 
regeneration processes. For all $\theta$, the outgoing flux of astrophysical 
neutrinos exceeds the background of atmospheric neutrinos at 
$E_{\nu} \gtrsim 10^5$\,GeV. This cross-over determines the energy threshold 
for detection of diffuse neutrino fluxes in UNTs. Results for Protheroe 
spectrum are in a qualitative agreement with ones published in \cite{kwi} (see
Fig.\,8 and Fig.\,9 there) where different models for neutrino interactions 
and $\tau$-lepton energy losses were used. 
The fractions of secondary $\nu_{\mu,e}$'s produced in regeneration
chains with respect to the total $\nu_{\mu,e}$
flux emerging after propagation through the Earth (made of
of primary $\nu_{\mu,e}$'s $+$ secondaries) are 0.57, 
0.18, 0.06, 0.02 (spectrum \cite{proth}) and 
0.18, 0.06, 0.02, 0.01 (spectrum \cite{mpr}) for nadir angles $\theta=$6$^o$, 
60$^o$, 80$^o$, and 85$^o$, respectively. These numbers are larger than 
corresponding fractions for power-law spectra 
$\phi_{\nu} \propto E_{\nu}^{-1}$ and $\phi_{\nu} \propto E_{\nu}^{-2}$ 
reported in \cite{dutta2}~\footnote{Authors do not provide numbers, 
nevertheless one can estimate them from Fig. 1 in \cite{dutta2}.}. On the 
other hand, the comparison of the results on attenuated neutrino fluxes 
obtained by our algorithm with the ones obtained in \cite{dutta2} for 
power-law spectra shows a reasonable agreement (Fig.~\ref{f11}). Hence, we
can conclude that larger fractions of secondaries result from the fact that 
MPR and Protheroe spectra are harder compared to ones considered in 
\cite{dutta2}.

\section{Tau leptons: detection signatures}
\label{sec:sign}

\subsection{Tau lepton identification through energy loss properties}
\label{sec:idenener}

\begin{figure}[htb]
\begin{center}
\includegraphics[height=12.0cm]{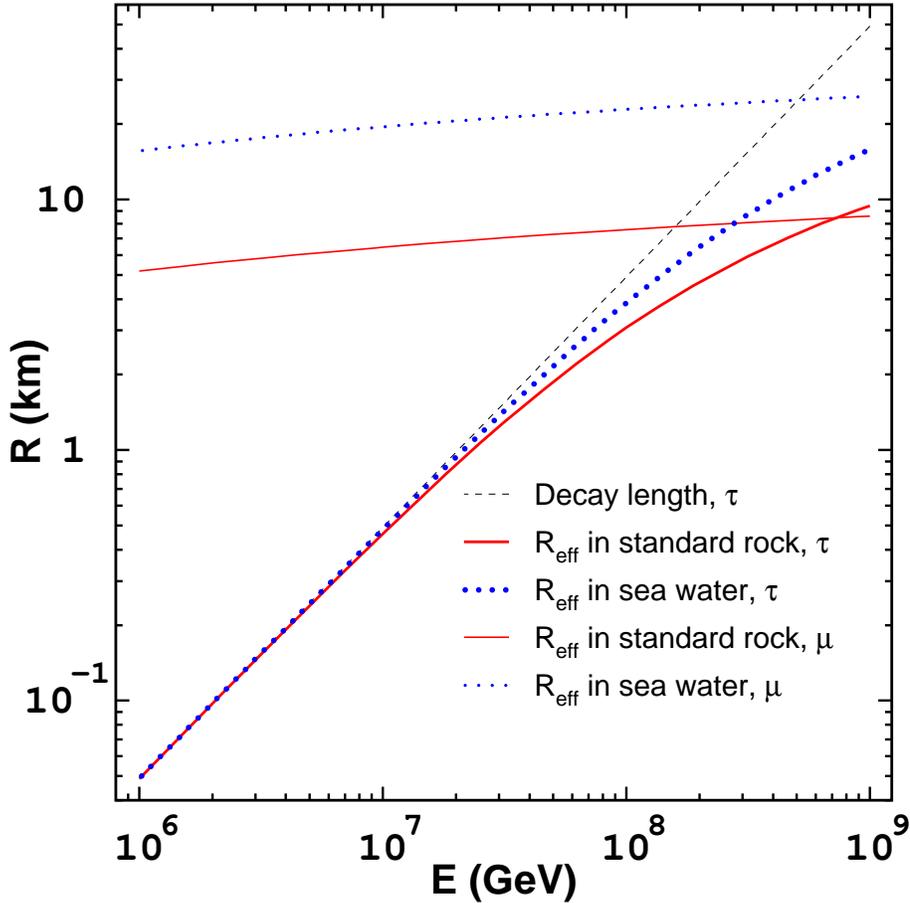}
\end{center}
\caption{\label{f12}
Mean ranges for muons and $\tau$-leptons in standard rock and sea water.
}
\end{figure}

\begin{figure}[htb]
\begin{center}
\includegraphics[height=13.2cm]{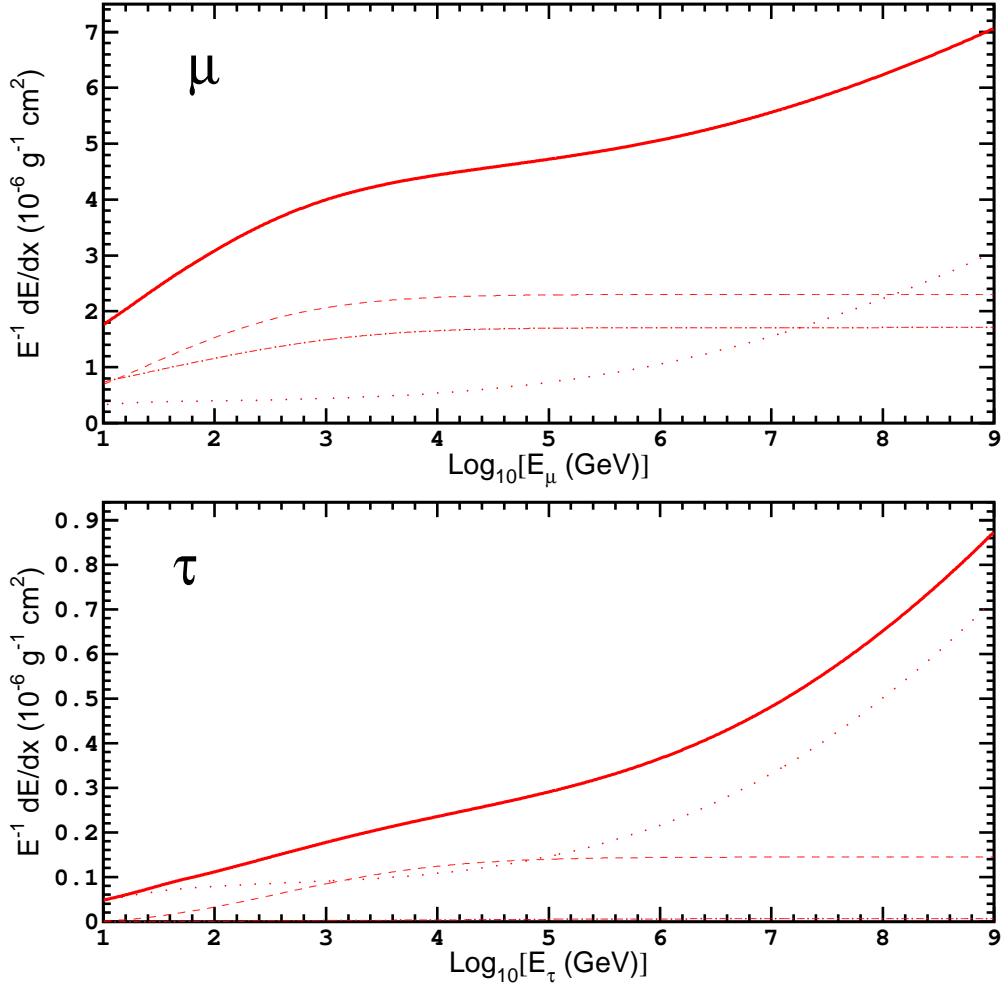}
\end{center}
\caption{\label{f13}
Muon (upper plot) and $\tau$-lepton (lower plot) energy losses in standard 
rock due to bremsstrahlung (dash-dotted lines), direct $e^{+}e^{-}$-pair 
production (dashed lines), photonuclear interaction (dotted lines). Thick 
solid lines stand for total radiative energy losses. Bremsstrahlung for 
$\tau$-lepton is strongly suppressed due to $m^{-2}$ factor, therefore the 
bremsstrahlung curve almost coincides with the $x$-axis.}
\end{figure}

At energies larger than $E_{\tau}\approx$\,2$\cdot$10$^6$ GeV 
$\tau$-lepton ranges 
become comparable to the typical linear dimensions of operating and proposed 
UNTs (see Fig.~\ref{f12}) and $\tau$ tracks can be reconstructed. As a first
guess, one could try to distinguish a $\tau$-lepton track from a muon one by 
means of differences in energy losses. The larger $\tau$
mass affects bremsstrahlung, $e^{+}e^{-}$-pair production and photonuclear 
processes differently respect to muons. As a result, $e^{+}e^{-}$-pair 
production dominates muon energy losses up to $E_{\mu} \approx$\,10$^{8}$\,GeV
while $\tau$-lepton photonuclear interaction plays the main role above 
$E_{\tau} \approx$\,10$^{5}$\,GeV (see Fig.~\ref{f13}) compared to other 
radiative processes. Hence, since probabilities of shower production for muons
and $\tau$-leptons at a given energy are different, one could hope to select 
$\tau$-lepton events studying the distribution of showers produced in 
photonuclear or electromagnetic interactions along tracks even if the detector
resolution is not so good to discriminate hadronic and electromagnetic showers.


\begin{figure}[htb]
\begin{center}
\includegraphics[height=14.1cm]{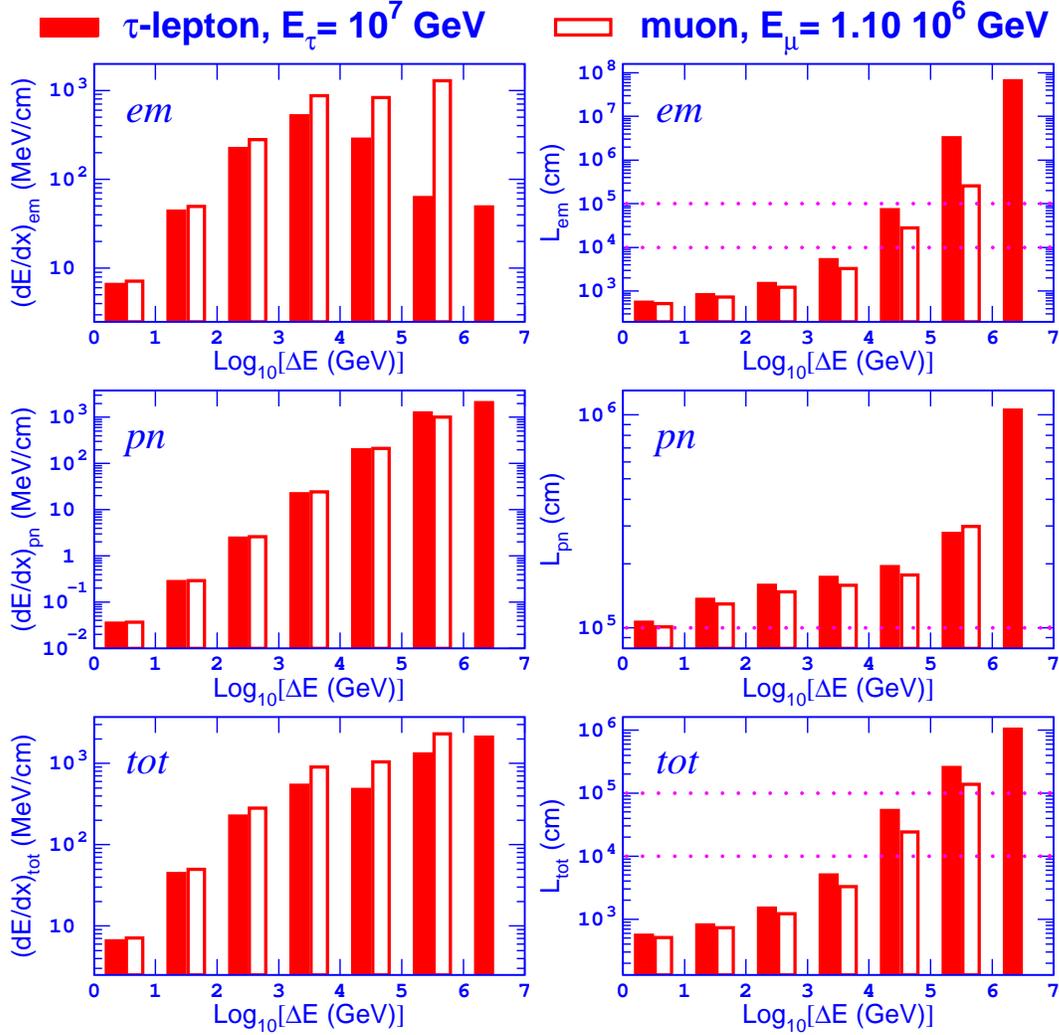}
\end{center}
\caption{\label{f14}
Left column: energy losses in sea water due to electromagnetic interactions 
(upper panel), photonuclear interactions (middle panel) and all interactions 
(lower panel) for energy transferred to secondaries belonging to different
ranges  ($10^{0} GeV \le \Delta E \le 10^{1} GeV$, 
$10^{1} GeV \le \Delta E \le 10^{2} GeV$, ..., 
$10^{6} GeV \le \Delta E \le 10^{7} GeV$). Filled boxes correspond
to a $\tau$-lepton with $E_{\tau}$ = 10$^{7}$ GeV, empty ones to a muon with
energy $E_{\mu}$ = $1.1\,10^{6}$ GeV. Right column: Mean interaction length of
muon/$\tau$-lepton in sea water for different ranges of energy transfered to 
secondaries. The horizontal lines indicate 100\,m and 1\,km which are 
the typical linear sizes of existing and planned UNTs.
}
\end{figure}

\begin{figure}[htb]
\begin{center}
\includegraphics[height=14.1cm]{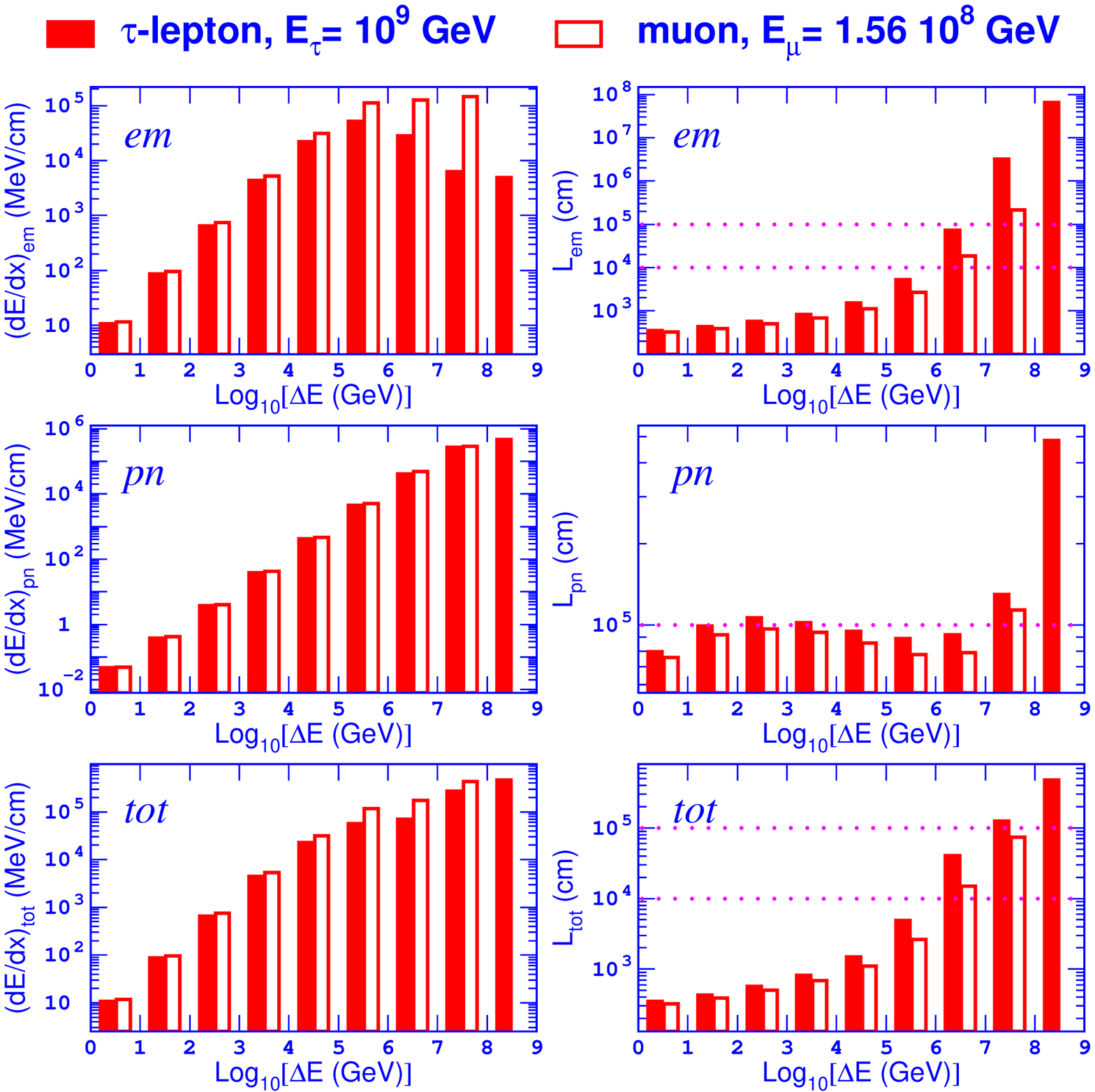}
\end{center}
\caption{\label{f15}
The same as in Fig.~\ref{f14} for a 10$^{9}$ GeV $\tau$-lepton and 
a $1.56 \cdot 10^8$ GeV muon.  
}
\end{figure}

Nevertheless, the results of the calculation that we have performed for
$\tau$-leptons with energies in the range 
10$^6$\,GeV$\le E_{\tau} \le$10$^9$\,GeV indicate that such a method works 
badly because the largest differences between shapes of $\tau$-lepton and muon 
cross-sections lies in a range of relatively large $v=\Delta E/E$, where 
interaction lengths exceed the detector size. The numerical data on 
$\tau$-leptons of energies 10$^{7}$ GeV and 10$^{9}$ GeV and muons of lower 
energies (1.10$\cdot$10$^{6}$ and 1.56$\cdot$10$^{8}$, respectively) are 
presented in Fig.~\ref{f14} and Fig.~\ref{f15}, which show values of 
energy losses and radiation lengths for different ranges of energy transfered 
to secondaries ($10^{0} GeV \le \Delta E \le 10^{1} GeV$, 
$10^{1} GeV \le \Delta E \le 10^{2} GeV$, ..., 
$10^{6} GeV \le \Delta E \le 10^{7} GeV$). Muon energies 
were chosen so that total muon energy losses are equal to total
energy losses of considered $\tau$-leptons. Data on electromagnetic processes 
(bremsstrahlung $+$ $e^{+}e^{-}$-pair production), photonuclear interaction 
and all interactions are presented separately. One can see that 'partial'
energy losses and radiation lengths of $\tau$-leptons and muons of lower 
energies differ remarkably only for large transferred energies when radiation 
length exceeds 100 - 1000 m, typical linear dimensions of existing and 
planned detectors. Thus, there are too few high energy showers inside the 
detector sensitive volume to provide enough statistics to distinguish
$\tau$-lepton and muon tracks. Even in the hypothesis that the detector 
capability is so good to distinguish electromagnetic showers from hadronic 
ones, it is very difficult and, most probably, impossible to distinguish 
a $\tau$-lepton track of a certain energy with respect to a muon track of 
$\sim$6$\div$11 times lower energy. This difficulty also concerns 
\mbox{1 km$^{3}$-scale} UNTs. Of course, the observation of a very high energy
shower with reconstructed energy so large to be inconsistent with the 
energy reconstructed using the rest of the track could be an indication in 
favor of a $\tau$-lepton. Nevertheless, the probability of such a shower 
occurrence is too small that this cannot be considered a standard reliable 
method. 

\subsection{Topological signatures}

As a matter of fact, signatures such as 'double bang' (DB) or 'lollipop' 
events proposed in \cite{db} seem to be smoking guns to recognize 
$\tau$-lepton events in UNTs. DB event features are an hadronic cascade at the 
$\nu\,N$ CC interaction vertex and an hadronic or electromagnetic cascade
corresponding to $\tau$-lepton decay. The $\tau$-lepton track connects both 
cascades~\footnote{One should note that at energies where the $\tau$-lepton 
track connecting the 2 cascades can be reconstructed in a UNT 
($E_{\tau} \gtrsim 1$\,PeV), the $\tau$ is not a minimum ionizing track, as 
instead it is assumed in \cite{db}. In the example considered in \cite{db} (a 
1.8 PeV $\tau$-lepton) about 5.5$\cdot$10$^{8}$ \u{C}erenkov photons in the 
wavelength range 350\,nm $\le \lambda \le$ 600\,nm are emitted by a 90\,m long
track. This number is $\sim 300$ times larger than for a minimum ionizing
track. Nonetheless it is still several orders of magnitude lower than the 
number of photons from both cascades ($\approx$4$\cdot$10$^{11}$ and 
$\approx$10$^{11}$ from first and second cascades, respectively), thus the 
signature of DB remains unchanged with this correction.}. A lollipop event 
consists of a partially contained $\tau$-lepton track with the first cascade 
at the $\nu$ interaction vertex outside of the UNT sensitive volume and with 
the second cascade produced by $\tau$-lepton decay contained in the volume. On
the other hand, the inverse pattern with only the first cascade visible can 
not be distinguished from $\nu_{\mu}\,N \to \mu\,N$ interaction followed by 
the muon track. Both DB and lollipop signatures provide a clear and 
background-free evidence of $\nu_{\tau}$ detection. 
As a matter of fact, for $\tau$ decaying into electronic and hadronic channels
there is no muon track following the second cascade. Besides, DB events do not
contain an in-coming track before the first cascade. In principle, an up-going
muon generated in a $\nu_{\mu}$ CC interaction could mimic such a signature if
it would produce a shower that takes all the muon energy inside the 
instrumented volume. But the probability to have such an event for a muon 
track segment of the order of 1 km is less than 5$\cdot$10$^{-4}$ in water for
energies larger than 1 PeV. This value corresponds to the probability that a 
muon looses more than 99\% of its energy in one interaction. This number can 
be considered as an upper limit on the signal to noise ratio, where the signal
are DB/lollipop events and the noise are up-going $\nu$-induced muons. In 
fact, astrophysical $\nu_{\mu}$ fluxes are lower or at most equal to expected 
astrophysical $\nu_{\tau}$ ones after propagation in the Earth and atmospheric
neutrino fluxes are even lower above 1 PeV (see Fig.~\ref{f9} and 
Fig.~\ref{f10}). Another background source could be due to atmospheric muons 
that are dominated by prompt muons in the considered energy range. DB events 
cannot be reproduced since in any case the muon track before the first cascade
would be visible. Nevertheless, a muon can mimic a lollipop event if it enters
the detector instrumented volume and undergoes an interaction with such a 
large energy transfer that the muon stops after shower production. We estimate
a rate of $\sim 50$ km$^{-3}$ yr$^{-1}$ considering the probability upper 
limit of $5 \cdot 10^{-4}$ (corresponding to more than 99\% muon energy loss)
and the prompt muon fluxes published in \cite{gelmini,sineg}. However, this is
an upper limit since it is not possible to compute exactly the probability 
that a muon looses 100\% of its energy since cross section parameterizations 
close to $v= \Delta E/E = 1$ do not describe perfectly well the real cross 
sections. 
Of course more quantitative conclusions on possible backgrounds are needed 
that should be based on detailed simulations including detector response and 
reconstruction algorithms. At least qualitatively, DB and lollipop events  
are potentially ``background free'', as it has been stated in the paper that 
proposed the existence of these topologies \cite{db}.

In 17\% of the cases, that is for the $\tau$ decay channel into a muon (see 
eq.~\ref{decay3}), the second cascade is absent. In this case the 
$\tau$-lepton track is followed by a muon track without any shower. This 
topology seems to be not-recognizable as a $\tau$-lepton signature in a UNT, 
although the amount of light produced by showers along the $\tau$ track 
(hereafter 'brightness')  differs essentially from that along a muon track of 
the same energy. As an example, let us consider a $\tau$-lepton of energy 
$E_{\tau}$=\,10$^{7}$\,GeV that decays into a muon. For a simple estimation, 
we assume that the muon takes 1/3 of the $\tau$ energy, thus 
$E_{\mu} \approx$\,3.3$\cdot$10$^{6}$\,GeV. From  results in 
Sec.~\ref{sec:idenener} (see Fig.~\ref{f14}) we know that the brightness of 
a 10$^{7}$\,GeV $\tau$ track is similar to that of a $\mu$ track with 
$E_{\mu} = 1.1 \cdot 10^{6}$\,GeV. After $\tau$-lepton decay, the track 
brightness increases by about a factor of 
$\approx$3.3$\cdot$10$^{6}$\,GeV$/$1.1$\cdot$10$^{6}$\,GeV$=$3. This factor 
depends on $\tau$-lepton energy and in the range 
10$^6$\,GeV\,$\le E_{\tau} \le$\,10$^9$\,GeV it varies between 2 and 4. To 
recognize such a signature detector energy reconstruction should be remarkably
better than this factor, which is not always the case for neutrino telescopes
(see for instance Fig.\,4 from \cite{enerres}). Thus, when computing DB or/and
lollipop event rates in a UNT, results must be reduced by a factor 
$B_{-\mu}$=\,0.83 to exclude the non-muonic branching ratio in $\tau$-lepton 
decay.

Estimates of DB rates in km$^3$ UNTs have already been presented 
in~\cite{athar} for down-going $\nu_{\tau}$'s, while events from the lower 
hemisphere were not considered. Calculations of up-going $\nu_{\tau}$ 
propagation were done in~\cite{dutta1}. Authors give a qualitative conclusion
that though UHE/EHE $\nu_{\tau}$'s are not absorbed by the Earth, their 
energies decrease; hence the expected amount of $\tau$ events in the DB energy
range is low. Calculations in \cite{beacom1} performed for a neutrino 
flux $E^{2}_{\nu_{\mu}}\,dN_{\nu_{\mu}}\,/\,dE_{\nu_{\mu}}=$\,10$^{-7}$\,GeV\,cm$^{-2}$\,s$^{-1}$ 
assuming $\nu$ oscillations but without accounting for neutrino absorption in 
the Earth result in a rate of $\approx$0.5\,yr$^{-1}$ both for DB and lollipop
events in an IceCube-like km$^3$ detector \cite{icecube}. The measurement of 
the ratio between muon and shower events that may be an indirect signature of 
$\nu_{\tau}$ appearance was proposed in \cite{dutta1,beacom1} as an 
alternative indirect way to detect $\tau$-neutrinos. Here we present results 
of a calculation of the DB rate for both the upper and lower hemispheres in a 
km$^3$ scale UNTs using results reported in Sec.~\ref{sec:results} for the 
spectra in \cite{proth,mpr} that were not considered in 
\cite{dutta1,athar,beacom1}.
The rate of totally contained DB events in a UNT is given by:
{\normalsize
\begin{equation}
\label{numberevents}
n_{DB}=2\pi\,\rho\,N_A\,B_{-\mu}\!\!\!\int_{-1}^{0(1)}\!\!\int^{\infty}_{E_{min}}\!\!V_{eff}(E_{\nu_{\tau}},\theta)\,I(E_{\nu_{\tau}},\theta)\,\sigma^{CC}(E_{\nu_{\tau}})\,dE_{\nu_{\tau}}d(\cos\theta)
\label{eq:rate}
\end{equation}}

\noindent
Here $N_A$ is the Avogadro number, $\rho$ is the medium density (we use
$\rho=$\,1\,g\,cm$^{-3}$ which is close to sea water/ice density), 
$I(E_{\nu_{\tau}},\theta)$ is the differential $\nu_{\tau}$ flux. The Earth 
shadowing effect is accounted for in the case of the lower hemisphere (for
which the upper limit of the integral is 0, while $\cos\theta=1$ is used to 
compute the number of events for the whole sphere). The effective volume is: 
\begin{equation}
V_{eff}=S_{p}(\theta)\,(L-R_{\tau}(E_{\nu_{\tau}}))\, , 
\label{eq:veff}
\end{equation}
where $S_p$ is the projected area for tracks generated isotropically in 
azimuth for a fixed $\theta$ direction of the incident neutrino on a 
parallelepiped. The parallelepiped has the dimensions of an IceCube-like 
detector \cite{icecube} ($1\!\times\!1\!\times\!1$\,km$^3$) or a NEMO-like one
\cite{nemo} ($1.4\!\times\!1.4\!\times\!0.6$\,km$^3$). In eq.~\ref{eq:veff}, 
$L$ is the geometrical distance between entry and exit point from the 
parallelepiped, $R_{\tau}$ is the $\tau$-lepton range, $\sigma^{CC}$ is total 
CC $\nu$ cross-section. $E_{min}\!=\!2\cdot10^6$\,GeV corresponds to a 
$\tau$-lepton range of $R^{min}_{\tau}\!=\!70$\,m. 

\begin{table}[htb]
\caption{\label{tablerate} Number of DB events in km$^3$ detector per year.}
\begin{center}
\begin{tabular}{ccc}
\hline
\hline
Spectrum \rule{0pt}{4.0mm} \rule[-1mm]{0pt}{2.5mm} & IceCube-like ($N_{-2\pi}\,/\,N_{2\pi}\,/\,N_{4\pi}$) & NEMO-like ($N_{-2\pi}\,/\,N_{2\pi}\,/\,N_{4\pi}$)\\
\hline
$Protheroe$ \rule{0pt}{4.0mm} \rule[-1mm]{0pt}{2.5mm} & $0.6\,\,\,\,/\,\,\,\,1.3\,\,\,\,/\,\,\,\,1.9$      & $0.8\,\,\,\,/\,\,\,\,1.7\,\,\,\,/\,\,\,\,2.5$ \\
$MPR$  \rule{0pt}{4.0mm} \rule[-1.5mm]{0pt}{2.5mm} &  $0.8\,\,\,\,/\,\,\,\,1.9\,\,\,\,/\,\,\,\,2.7$      & $1.2\,\,\,\,/\,\,\,\,2.6\,\,\,\,/\,\,\,\,3.8$ \\
\hline
\hline
\end{tabular}
\end{center}
\end{table}

Table~\ref{tablerate} shows DB rates for the lower, upper and for both 
hemispheres. Values for upper hemisphere are 3$\div$6 times lower compared to 
\cite{athar} since more optimistic predictions for diffuse neutrino fluxes 
\cite{SS} are used there. At the same time, our figures for DB rates in 
IceCube-like UNT are 4$\div$5 higher compared to results reported in 
\cite{beacom1}. 
There are at least four factors that contribute to this
difference:
\begin{itemize}
\item
different $\nu$ flux models are used;
\item
minimum $\tau$ range is $R^{min}_{\tau}\!=\!70$\,m in our calculations while 
in \cite{beacom1} it is equal to $R^{min}_{\tau}\!=\!200\div 400$\,m. Such a 
large minimum range was chosen according to the qualitative requirement to 
reconstruct the $\tau$-lepton track connecting the two cascades. This
difference accounts for a factor of $\sim$4 in the case of 
an $E^{-2}$
$\nu$ flux between $10^6$ and $10^7$ GeV. 
Since horizontal 
photomultiplier spacing in IceCube detector is $\sim$125\,m, but the vertical 
spacing is 17\,m \cite{icecube}, we believe that this requirement could be too
restrictive. The proper choice of the minimum range should indeed be done 
quantitatively using a full simulation of the detector including 
reconstruction procedures;
\item
for the lower hemisphere we accounted for $\tau$-neutrino energy losses in the
Earth. This losses can be such that events do not fall anymore above the DB 
`energy threshold' (that figures for lower hemisphere in 
Table~\ref{tablerate} are about a factor of $\approx$2 lower compared to the 
upper hemisphere ones). These losses were not accounted for in \cite{beacom1};
\item
in contrast to \cite{beacom1} we reduce all figures for rates by a factor
$B_{-\mu}$=\,0.83 to exclude the muonic mode of $\tau$-lepton decay.
\end{itemize}

\section{Conclusions}
\label{sec:conclusions}

Considering the AGN diffuse neutrino flux models in \cite{proth} and the 
upper limit on optically thick $\nu$ sources in \cite{mpr} we have found 
that in a km$^3$ UNT one can expect 2$\div$4 DB events yr$^{-1}$~\footnote{The
rate of lollipop events is expected to be slightly larger than that of DB 
events. In fact, for lollipops the second cascade is not required to be inside
an UNT, while for DB both should be contained. Hence, the detection 
probability for lollipop events is essentially larger than for DB at 
$E_{\tau} \gtrsim$\,10\,PeV \cite{beacom1}, but due to the steepness of
neutrino spectra the difference in rates is not very large. 
Nevertheless, an estimate would require a more detailed detector
simulation than what has been done in this work.}. 
This result depends on the assumed neutrino spectrum and on
the minimum $\tau$-lepton range that can be reconstructed in the detector, 
hence on its geometry and reconstruction algorithms. Hence, these values
should be considered as indicative, while full simulations for specific 
neutrino telescopes should be performed. Moreover, we have investigated the 
possibility of identification of $\tau$-lepton from muons by studying the 
distribution of showers produced by electromagnetic and photonuclear 
interactions along the track. This seems to be very difficult. Detection of DB
or lollipop events are the cleanest
signatures even though rates seem to be quite small. An alternative indirect 
way to identify $\tau$'s consists in measuring the ratio between shower and 
track events. In any case, simulations of $\tau$-neutrino propagation through 
the Earth should be performed including regeneration processes and 
$\tau$-lepton energy losses. For this, in particular, one needs to account for
corrections to the soft part of the photonuclear interaction which increases 
the total $\tau$-lepton energy losses by $\approx$20$\div$30\% in UHE/EHE 
range.

\section*{Acknowledgments}

We thank P.~Lipari for providing the original $\nu$ cross-section generator.

\appendix

\section{Appendix: photonuclear cross-section}
\label{app:PN}

The differential cross-section for photonuclear interaction used in this work 
has been published in \cite{bugaev1}. It is rewritten below in the form 
adopted in \cite{mum} (Appendix A). In the formula the following values are 
used:
$\alpha$ = 7.297353$\times$10$^{-3}$ -- fine structure constant;
$A \equiv$ atomic weight;
$m_{l} = m_{\mu} =$\,0.1057\,GeV for muon and 
$m_{l} = m_{\tau} =$\,1.777\,GeV for $\tau$-lepton; 
$m_N \equiv$ nucleon mass;
$\pi$ = 3.141593.

{\normalsize
\begin{table}
\caption{
{\normalsize \label{ak}
Coefficients $a_k$ for $\frac{d\sigma^{pn}_{hard}}{d v} = \frac{1}{v}\sum^{7}_{k=0}\,a_{k}\log_{10}^kv$ 
in the formula for photonuclear cross-section. The upper figures stand for 
muons, lower ones does for $\tau$-leptons.
}}
\begin{center}
\begin{tabular}{ccccc}
\hline
\hline
E ($GeV$) \rule[-2mm]{0pt}{6mm} & {\normalsize $a_0$} & {\normalsize $a_1$} & {\normalsize $a_2$} & {\normalsize $a_3$}\\
\hline
$10^3$ \rule{0pt}{4mm} & 7.174409$\cdot$10$^{-4}$  & -0.2436045 & -0.2942209 & -0.1658391 \\[-3.0mm]
\strut \rule[-2mm]{0pt}{2mm} & -1.269205$\cdot$10$^{-4}$ &  -0.01563032 & 0.04693954 &  0.05338546 \\
\hline
$10^4$ \rule{0pt}{4mm} & 1.7132$\cdot$10$^{-3}$  & -0.5756682 & -0.68615 & -0.3825223 \\[-3.0mm]
\strut \rule[-2mm]{0pt}{2mm}  & -2.843877$\cdot$10$^{-4}$ & -0.03589573 & 0.1162945 & 0.130975\\
\hline
$10^5$ \rule{0pt}{4mm} &  4.082304$\cdot$10$^{-3}$ & -1.553973 & -2.004218 & -1.207777\\[-3.0mm]
\strut \rule[-2mm]{0pt}{2mm} &  -5.761546$\cdot$10$^{-4}$ & -0.07768545 & 0.3064255 & 0.3410341\\
\hline
$10^6$ \rule{0pt}{4mm} & 8.628455$\cdot$10$^{-3}$ & -3.251305 &  -3.999623 &  -2.33175\\[-3.0mm]
\strut \rule[-2mm]{0pt}{2mm} & -1.195445$\cdot$10$^{-3}$ & -0.157375 & 0.7041273 & 0.7529364\\
\hline
$10^7$ \rule{0pt}{4mm} & 0.01244159 & -5.976818 & -6.855045 & -3.88775\\[-3.0mm]
\strut \rule[-2mm]{0pt}{2mm} & -1.317386$\cdot$10$^{-3}$ & -0.2720009 & 1.440518 & 1.425927\\
\hline
$10^8$ \rule{0pt}{4mm} & 0.02204591 & -9.495636 & -10.05705 & -5.636636\\[-3.0mm]
\strut \rule[-2mm]{0pt}{2mm} & -9.689228$\cdot$10$^{-15}$ & -0.4186136 & 2.533355 & 2.284968\\
\hline
$10^9$ \rule{0pt}{4mm} & 0.03228755 & -13.92918 & -14.37232 & -8.418409\\[-3.0mm]
\strut \rule[-2mm]{0pt}{2mm} & -6.4595$\cdot$10$^{-15}$ & -0.8045046 & 3.217832 & 2.5487\\
\hline
\hline
E ($GeV$) \rule[-2mm]{0pt}{6mm}  & {\normalsize $a_4$} & {\normalsize $a_5$} & {\normalsize $a_6$} & {\normalsize $a_7$}\\
\hline
$10^3$ \rule{0pt}{4mm} & -0.05227727 &  -9.328318$\cdot$10$^{-3}$ & -8.751909$\cdot$10$^{-4}$ & -3.343145$\cdot$10$^{-5}$\\[-3.0mm]
\strut \rule[-2mm]{0pt}{2mm} & 0.02240132 & 4.658909$\cdot$10$^{-3}$ & 4.822364$\cdot$10$^{-4}$  & 1.9837$\cdot$10$^{-5}$\\
\hline
$10^4$ \rule{0pt}{4mm} & -0.1196482 &  -0.02124577 & -1.987841$\cdot$10$^{-3}$ & -7.584046$\cdot$10$^{-5}$\\[-3.0mm]
\strut \rule[-2mm]{0pt}{2mm} & 0.05496 & 0.01146659 & 1.193018$\cdot$10$^{-3}$ & 4.940182$\cdot$10$^{-5}$ \\
\hline
$10^5$ \rule{0pt}{4mm} &  -0.4033373 & -0.07555636 & -7.399682$\cdot$10$^{-3}$ & -2.943396$\cdot$10$^{-4}$ \\[-3.0mm]
\strut \rule[-2mm]{0pt}{2mm} & 0.144945 & 0.03090286 & 3.302773$\cdot$10$^{-3}$ & 1.409573$\cdot$10$^{-4}$ \\
\hline
$10^6$ \rule{0pt}{4mm} &   -0.7614046 & -0.1402496 & -0.01354059 & -5.3155$\cdot$10$^{-4}$ \\[-3.0mm]
\strut \rule[-2mm]{0pt}{2mm} & 0.3119032 & 0.06514455 & 6.843364$\cdot$10$^{-3}$ & 2.877909$\cdot$10$^{-4}$ \\
\hline
$10^7$ \rule{0pt}{4mm} & -1.270677 & -0.2370768 & -0.02325118 &  -9.265136$\cdot$10$^{-4}$ \\[-3.0mm]
\strut \rule[-2mm]{0pt}{2mm} & 0.5576727 & 0.1109868 & 0.011191 & 4.544877$\cdot$10$^{-4}$ \\
\hline
$10^8$ \rule{0pt}{4mm} & -1.883845 & -0.3614146 & -0.03629659 & -1.473118$\cdot$10$^{-3}$ \\[-3.0mm]
\strut \rule[-2mm]{0pt}{2mm} & 0.8360727 & 0.1589677 & 0.015614 & 6.280818$\cdot$10$^{-4}$ \\
\hline
$10^9$ \rule{0pt}{4mm} & -2.948277 & -0.5819409 & -0.059275 & -2.419946$\cdot$10$^{-3}$  \\[-3.0mm]
\strut \rule[-2mm]{0pt}{2mm} & 0.8085682 & 0.1344223 & 0.01173827 & 4.281932$\cdot$10$^{-4}$ \\
\hline
\hline
\end{tabular}
\end{center}
\end{table}
}

{\normalsize
$$
\begin{array}{l}
\frac{d\sigma^{pn}}{d v} = \frac{\alpha}{8\pi}A\sigma_{\gamma N}\,
                 v\left\{\left(H(v)+\frac{2m_{l}^2}{m_2^2}\right)\ln\left(1+\frac{m_2^2}{t}\right)\right.\\[5mm]
           \qquad -\left.\frac{2m_{l}^2}{t}\left[1-\frac{0.25m_2^2}{t}
                 \ln\left(1+\frac{t}{m_2^2}\right)\right]
              +G(z)\left[H(v)\left(\ln\left(1+\frac{m_1^2}{t}\right)-\frac{m_1^2}{m_1^2+t}\right)\right.\right.\\[5mm]
      \qquad     +\left.\left.\frac{4m^2_{l}}{m^2_1}\ln\left(1+\frac{m^2_1}{t}\right)
-\frac{2m_{l}^2}{t}\left(1-\frac{0.25m_1^2-t}{m_1^2+t}\right)\right]\right\}+A\frac{d\sigma^{pn}_{hard}}{d v},
\end{array}
$$
\[
H(v)=1-\frac{2}{v}+\frac{2}{v^2}, \qquad
\]
\begin{eqnarray*}
G(z) & = & \frac{9}{z}\left[\frac{1}{2}+\frac{(1+z)e^{-z}-1}{z^2}\right] \qquad (Z\ne1), \\
G(z) & = & 3 \qquad (Z=1),
\end{eqnarray*}
\[
z     = 0.00282A^{1/3}\sigma_{\gamma N},\quad
t     = \frac{m_{l}^2 v^2}{1-v}, \quad
m_1^2 = 0.54\;{\rm GeV}^2, \quad
m_2^2 = 1.80\;{\rm GeV}^2
\]
with total cross-section for absorption of a real photon of energy 
$\nu=s/2m_N=vE$ by a nucleon, $\sigma_{\gamma N}$, parameterized according to 
\cite{pbb1,pbb2}
\[
\sigma_{\gamma N}=[114.3+1.647 \ln^{2}(0.0213\,\nu)]\,\mu b
\]
and hard part of cross-section described by polynomial parameterization
\[
\frac{d\sigma^{pn}_{hard}}{d v} = \frac{1}{v}\sum^{7}_{k=0}\,a_{k}\log_{10}^kv
\]
}

\noindent
(coefficients $a_k$ are given in Table \ref{ak}). Parameterization for
$\frac{d\sigma^{pn}_{hard}}{d v}$ is valid in the range $10^{-6}\le v \le 1$,
$10^3 \le E \le 10^9$\,$GeV$, in MUM\,1.5 $\frac{d\sigma^{n}_{hard}}{d v}$    
is assigned to 0 for $E=$\,10$^2$\,$GeV$ and $v=10^{-7}$, values between 
$v=10^{-7}$ and $v=10^{-6}$, as well as between $E=$\,10$^2$\,GeV and
$E=$\,10$^3$\,GeV are computed by interpolation. Values for energies between 
ones that are given in the Table~\ref{ak} are computed by interpolation, as 
well.


\newpage

\begin{thebibliography}{99}

\bibitem{amanda}
AMANDA Collaboration:
P.~Desiati et al., astro-ph/0306536; 
see URL $<$http://amanda.uci.edu/$>$.

\bibitem{antares}
ANTARES Collaboration: 
E.~Aslanides et al., astro-ph/9907432; 
T.~Montaruli et al., in: Proceedings of 28th International Cosmic Ray 
Conference, Tsukuba, 2003, p.\,1357. Available from 
$<$arXiv:physics/0306057$>$; 
see URL $<$http://antares.in2p3.fr/$>$.

\bibitem{baikal}
Baikal Collaboration:
R.~Wischnewski et al., in: Proceedings of 28th International Cosmic Ray 
Conference, Tsukuba, 2003. Available from $<$arXiv:astro-ph/0305302$>$; 
see URL $<$http://nt200.da.ru/$>$.

\bibitem{icecube}
IceCube Collaboration:
J.~Ahrens et al., astro-ph/0305196; see URL $<$http://icecube.wisc.edu/$>$.

\bibitem{nemo}
NEMO Collaboration:
G.~Riccobene et al., in: Proceedings of second Workshop on Methodical 
Aspects of Underwater/Underice Neutrino Telescopes, Hamburg, 2001, p.\,61; 
see URL $<$http://nemoweb.lns.infn.it/$>$.

\bibitem{nestor}
NESTOR Collaboration:
S.~E.~Tzamarias  et al., NIM A502 (2003) 150; see URL
$<$http://www.nestor.org.gr/$>$.

\bibitem{osc1} 
Super-Kamiokande Collaboration:
S.~Fukuda et al., Phys. Rev. Lett. 85 (2000) 3999. Available from
$<$arXiv:hep-ex/0009001$>$. 

\bibitem{osc2}
MACRO Collaboration:
M.~Ambrosio et al., Phys. Lett. B517 (2001) 59. Available from
$<$arXiv:hep-ex/0106049$>$.

\bibitem{osc3}
K2K Collaboration:
M.~H.~Ahn et al., Phys. Rev. Lett. 90 (2003) 041801. Available 
from $<$arXiv:hep-ex/0212007$>$. 

\bibitem{notau1} 
G.~Barenboin, C.~Quigg, Phys. Rev. D67 (2003) 073024. Available 
from $<$arXiv:hep-ph/0301220$>$. 

\bibitem{notau2} 
J.~F.~Beacom, Phys. Rev. Lett. 90 (2003) 181301. Available from 
$<$arXiv:hep-ph/0211305$>$. 

\bibitem{PDG}
Particle Data Group:
K.~Hagiwara et al., Phys. Rev. D66 (2002) 010001. 
Available from $<$http://pdg.lbl.gov/$>$.

\bibitem{kwi}
J.~Kwiecinski, A.~D.~Martin,  A.~M.~Stasto, Phys. Rev. D59 (1999) 
093002. Available from $<$arXiv:astro-ph/9812262$>$. 

\bibitem{dutta1}
S.~I.~Dutta,  M.~H.~Reno, I.~Sarcevic, Phys. Rev. D62 (2000) 123001. 
Available from $<$arXiv:hep-ph/0005310$>$. 

\bibitem{beacom}
J.~F.~Beacom, P.~Crotty, E.~W.~Kolb, Phys. Rev. D66 (2002) 021302(R). 
Available from $<$arXiv:astro-ph/0111482$>$.

\bibitem{dutta2}
S.~I.~Dutta, M.~H.~Reno, I.~Sarcevic, Phys. Rev. D66 (2002) 077302. 
Available from $<$arXiv:hep-ph/0207344$>$. 

\bibitem{yosh}
S.~Yoshida, R.~Ishibashi, H.~Miyamoto, astro-ph/0312078.
 
\bibitem{bottai}
S.~Bottai, S.~Giurgola, Astropart. Phys. 18 (2003) 539. Available 
from $<$arXiv:astro-ph/0205325$>$.

\bibitem{giesel}
K.~Giesel, J.-H.Jureit, E.~Reya, Astropart. Phys. 20 (2003) 335. 
Available from $<$arXiv:astro-ph/0303252$>$. 

\bibitem{sar1}
J.~Jones et al, hep-ph/0308042. 

\bibitem{db} J.~G.~Learned, S.~Pakvasa, Astropart. Phys. 3
(1995) 267. Available from $<$arXiv:hep-ph/9405296$>$. 

\bibitem{athar}
H.~Athar, G.~Parente, E.~Zas, Phys. Rev. D62 (2000) 093010. 
Available from $<$arXiv:hep-ph/0006123$>$.

\bibitem{dutta3}
S.~I.~Dutta, M.~H.~Reno, I.~Sarcevic, Phys. Rev. D64 (2001) 113015. 
Available from $<$arXiv:hep-ph/0104275$>$. 

\bibitem{beacom1}
J.~F.~Beacom et al., Phys. Rev. D68 (2003) 093005. Available from
$<$arXiv:hep-ph/0307025$>$.

\bibitem{ourtau1}
E.~Bugaev, T.~Montaruli, I.~Sokalski, in: Proceedings of 28th International 
Cosmic Ray Conference, Tsukuba, 2003, p.\,1381. Available from 
$<$arXiv:astro-ph/0305284$>$. 

\bibitem{ourtau2}
E.~Bugaev T.~Montaruli, I.~Sokalski, astro-ph/0311086. 

\bibitem{proth}
R.~J.~Protheroe, in Accretion Phenomena and Related Outflows,
ed. Wickramasinghe et al. (Astronomical Society of the Pacific, San Francisco,
1997).

\bibitem{mpr}
K.~Mannheim,  R.~J.~Protheroe, J.~P.~Rachen, Phys. Rev. D63 (2001) 
023003. Available from $<$arXiv:astro-ph/9812398$>$. 

\bibitem{bugaev1}
E.~V.~Bugaev, Yu.~V.~Shlepin, Phys. Rev. D67 (2003) 034027. Available
from $<$arXiv:hep-ph/0203096$>$. 

\bibitem{pbb1} L.~B.~Bezrukov, E.~V.~Bugaev, Yad. Fiz. 32
(1980) 1636 [Sov. J.~Nucl. Phys. 32 (1980) 847].

\bibitem{pbb2} L.~B.~Bezrukov and E.~V.~Bugaev, Yad. Fiz. 33 
(1981) 1195 [Sov. J.~Nucl. Phys. 33 (1981) 635].

\bibitem{earth}
A.~Dziewonski, {\it Encyclopedia of Solid Earth Geophysics}, ed. by 
D.~E.~James (Van Nostrand Reinhold, New York, 1989), p.\,331, we used the 
formula from this work cited in R.~Gandhi et al., Astropart. Phys.
5 (1996) 81. Available from $<$arXiv:hep-ph/9512364$>$.

\bibitem{lip1}
P.~Lipari, M.~Lusignoli, F.~Sartogo, Phys. Rev. Lett. {\bf 74} (1995) 4384. 
Available from $<$arXiv:hep-ph/9411341$>$. 
 
\bibitem{LEPTO}
LEPTO package. Code and manual available from 
$<$http://www3.tsl.uu.se/thep/lepto$>$;
G.~Ingelman,  A.~Edin, J.~Rathsman, Comput. Phys. Comm. 101 (1997) 108. 
Available from $<$arXiv:hep-ph/9605286$>$.

\bibitem{hadronization}
B.~Andersson et al., Phys. Rep. 97 (1983) 31.

\bibitem{JETSET}
JETSET and PYTHIA packages. Available from 
$<$http://www.thep.lu.se/$\sim$torbjorn/Phythia.html$>$; 
T.~Sj\"ostrand, Comput. Phys. Comm. 82 (1994) 74.

\bibitem{CTEQ}
H.~L.~Lai et al., Phys. Rev. D51 (1995) 4763. Available from 
$<$arXiv:hep-ph/9410404$>$. 

\bibitem{PDFLIB}
H.~Plothow--Besch, Comput. Phys. Comm. 75 (1993) 396.

\bibitem{CTEQ4}
H.~L.~Lai et al., Phys. Rev. D55 (1997) 1280. Available from 
$<$arXiv:hep-ph/9606399$>$. 

\bibitem{CTEQ5}
H.~L.~Lai et al., Eur. Phys. J. C12 (2000) 375. Available from 
$<$arXiv:hep-ph/9903282$>$.

\bibitem{CTEQ6}
J.~Pumplin et al., JHEP 7 (2002) 12. Available from  
$<$arXiv:hep-ph/0201195$>$.

\bibitem{CTEQ4GAN}
R.~Gandhi et al., Phys. Rev. D58 (1998) 093009. Available from 
$<$arXiv:hep-ph/9807264$>$.

\bibitem{mum}
I.~Sokalski, E.~Bugaev, S.~Klimushin, Phys. Rev. D64 (2001) 074015. 
Available from $<$arXiv:hep-ph/0010322$>$. 

\bibitem{gotsman}
E.~Gotsman, E.~M.~Levin, U.~Maor, Eur. Phys. J. C5 (1998) 303. 
Available from $<$arXiv:hep-ph/9708275$>$.

\bibitem{martin}
A.~D.~Martin, M.~G.~Ryskin, A.~M.~Stasto, Eur. Phys. J. C7 (1999) 643. 
Available from $<$arXiv:hep-ph/9806212$>$.

\bibitem{mueller} 
A.~N.~Mueller, Nucl. Phys. B415 (1994) 373.

\bibitem{nikolaev}
N.~N.~Nikolaev, B.~G.~Zakharov, Z. Phys. C49 (1991) 607.

\bibitem{forshaw}
J.~R.~Farshaw,  G.~Kerley, G.~Shaw, Phys. Rev. D60 (1999) 074012. 
Available from $<$arXiv:hep-ph/9903341$>$. 

\bibitem{duttaloss}
S.~I.~Dutta et al., Phys. Rev. D63 (2001) 094020. Available from
$<$arXiv:hep-ph/0012350$>$.

\bibitem{brembb1} 
L.~B. Bezrukov, E.~V.~Bugaev, in: Proceedings of 17th International Cosmic 
Ray Conference, Paris, 1981, Vol.\,7, p.\,102. 

\bibitem{brembb2} 
Yu.~M.~Andreev, L.~B. Bezrukov, E.~V.~Bugaev, Yad. Fiz. 57 (1994) 2146 
[Phys. At. Nucl. 57 (1994) 2066].

\bibitem{pk1} 
S.~R.~Kelner, Yu.~D.~Kotov, Sov. J. Nucl. Phys. 7 (1968) 237. 

\bibitem{pk2} 
R.~ P.~Kokoulin, A.~ A.~Petrukhin, Acta Phys. Acad. Sci. Hung. 29, 
Suppl.~4, (1970) 277.

\bibitem{pk3} 
R.~P.~Kokoulin, A.~A.~Petrukhin, in: Proceedings of 12th International 
Cosmic Ray Conference, Hobart, 1971, Vol.\,6, p.\,A2436. 

\bibitem{pk4} 
S.~R.~Kelner, R.~P.~Kokoulin, A.~A.~Petrukhin, Yad. Fiz. 62 (1999) 2042
[Phys. Atom. Nucl. 62 (1999) 1894].

\bibitem{pk5} 
S.~R.~Kelner,  Moscow Engineering Physics Inst. Preprint No. 016-97, 1997. 

\bibitem{lohman}
W.~Lohmann, R.~Kopp, R.~Voss, CERN Report 85-03 (1985).

\bibitem{ion1}
R.~M.~Sternheimer, Phys. Rev. 103 (1956) 511.

\bibitem{ion2}
R.~M.~Sternheimer, R.~F.~Peierls, Phys. Rev. B3 (1971) 3681.

\bibitem{ion3}
R.~M.~Sternheimer, M.~J.~Berger, S.~M.~Seltzer, Atom. Data Nucl. Data Tabl. 
30 (1984) 261.

\bibitem{HERA1} 
M.~Derrick et al., Phys. Lett. B316 (1993) 412; 

\bibitem{HERA2} 
M.~Derrick et al., Z. Phys. C65 (1995) 379; 

\bibitem{HERA3} 
M.~Derrick et al., Z. Phys. C72 (1996) 399. Available from 
$<$arXiv:hep-ex/9607002$>$. 

\bibitem{HERA4} 
I.~Abt et al., Nucl. Phys. B407 (1993) 515; 

\bibitem{HERA5} 
T.~Ahmed et al., Nucl. Phys B439 (1995) 471. Available from 
$<$arXiv:hep-ex/9503001$>$; 

\bibitem{HERA6} 
H.~Aid et al., Nucl. Phys B470 (1996) 3. Available from
$<$arXiv:hep-ex/9603004$>$.

\bibitem{tauola}
S.~Jadach, J.~ H.~Kuhn, Z.~Was, Comput. Phys. Commun. 64 (1991) 275.
 
\bibitem{SS}
F.~W.~Stecker, M.~H.~Salamon, Space Sci. Rev. 75 (1996) 341. 
Available from $<$arXiv:astro-ph/9501064$>$.

\bibitem{amanda1a}
J.~Ahrens et al., Phys. Rev. Lett. 90 (2003) 251101. Available
from $<$arXiv:astro-ph/0303218$>$.

\bibitem{amanda1b}
J.~Ahrens et al., astro-ph/0306209.

\bibitem{baikal1}
V.~Aynutdinov et al., astro-ph/0305302. 

\bibitem{naumov1}
V.~A.~Naumov, in: Proceedings of second Workshop on Methodical Aspects of
Underwater/Underground Telescopes, Hamburg, 2001, p.\,31. Available 
from $<$arXiv:hep-ph/0201310$>$.

\bibitem{gelmini}
G.~Gelmini, P.~Gondolo, G.~Varieschi, Phys. Rev. D67 (2003) 017301.
Available from $<$arXiV:hep-ph/0209111$>$. 

\bibitem{sineg}
T.~S.~Sinegovskaya, S.~I.~Sinegovsky, Phys. Rev. D63 (2001) 096004.
Available from $<$arXiV:hep-ph/0007234$>$. 

\bibitem{enerres}
J.~A.~Aguilar et al., astro-ph/0310130.

\end{thebibliography}
\end{document}